\documentclass[aps,pra,twocolumn,10pt]{revtex4-2}

\usepackage[english]{babel}
\usepackage{csquotes}
\usepackage{microtype} 

\usepackage{amsmath, amsthm, amssymb, dsfont}
\usepackage[normalem]{ulem}

\usepackage{graphicx}
\usepackage{booktabs, array}
\usepackage[abs]{overpic}
\usepackage{color}

\usepackage{enumerate}

\usepackage[colorlinks=true, citecolor=blue, linkcolor=blue, urlcolor=blue]{hyperref} 

\begin{document}

\title{
Floquet  Entanglement Generation in Parametrically Driven  Coupled Superconducting Qubits}

\author{Gustavo M. Meneses A.$^{1}$}
\author{Daniel Dominguez$^{1}$}
\author{María José Sánchez$^{1,2}$}
\affiliation{$^{1}$ Instituto Balseiro, Centro Atómico Bariloche, Av. Bustillo 9500, (8400) Bariloche (RN), Argentina.\\
$^{2}$ Instituto de Nanociencia y Nanotecnología (CNEA - CONICET), Nodo Bariloche, Av. Bustillo 9500, (8400) Bariloche (RN), Argentina. }

\begin{abstract}

We investigate the dynamical generation of entanglement in a system of two superconducting qubits coupled through a parametrically driven longitudinal interaction. Using Floquet theory and exact numerical simulations, we analyze the time evolution of the system initialized in a separable ground state. Our results reveal a nontrivial mechanism for entanglement generation, fundamentally distinct from the conventional resonant excitation to an entangled eigenstate.
We show that this mechanism emerges when two initially separable eigenstates are mixed by the periodic driving under multiphoton resonance conditions. Since the effect cannot be captured within a standard rotating-wave approximation, we employ generalized Van Vleck near-degenerate perturbation theory to derive an effective analytical description. Within this framework, we demonstrate that the sustained entanglement originates from the hybridization of the dominant Floquet states, namely those with the largest overlap with the initial ground state.
Furthermore, the degree of entanglement can be efficiently controlled through the driving amplitude. In particular, for specific amplitudes, the entanglement is fully suppressed. We term this phenomenon as coherent destruction of entanglement.

\end{abstract}
\maketitle

\section{Introduction}

High-fidelity entanglement generation and control are central requirements for scalable quantum computation, and superconducting circuits constitute one of the most advanced platforms for their implementation \cite{dicarlo2009demonstration, krantz2019quantum, kjaergaard2020superconducting, arute2019quantum}. While in multi-qubit architectures, parametric driving has been recently  employed to suppress static crosstalk and to realize fast, tunable gates in transmon \cite{mundada2019suppression, collodo2020implementation, jin2023fast, sung2021realization,ma2025parametric} and fluxonium \cite{weiss2022fast, zhang2024tunable, chakraborty2025tunable} devices among others, this technique also offers a rich paradigm for engineering quantum  dynamics. In this work, we theoretically investigate parametric driving as a primary mechanism for the direct generation of robust and sustained entanglement. We consider two qubits interacting via a longitudinal coupling that is harmonically modulated, i.e., $J(t) = J_0 + A \cos(\omega t)$. While dynamically generated entanglement has been previously explored in the context of strong qubit driving under static interactions \cite{sauer_2012, gramajo2017entanglement, song2020floquet, gramajo2021efficient, munyaev2021control, bastrakova2025dissipation, ilinskaya2026resonant}, its implementation via a driven coupler constitutes so far a less explored scenario.

We apply Floquet theory and numerical simulations to track the evolution of the system initialized in a separable ground state and identify the conditions under which the external drive induces entanglement, which is quantified via the  concurrence \cite{wootters1998entanglement}.
Our  analysis reveals two fundamentally distinct frequency-matching mechanisms to generate entanglement, for which the concurrence exhibits remarkable different patterns as a function of the the driving amplitude $A$ and the static coupling $J_0$.


The first mechanism, corresponds to population transfer from the separable ground state to  an excited entangled state at resonances, which we term “separable–entangled”  resonances (SER). The concurrence displays narrow maxima at these resonances, with a parameter dependence that closely resembles the amplitude-modulated interference patterns characteristic of Landau–Zener–Stückelberg–Majorana (LZSM) interferometry \cite{oliver2005mach, berns2008amplitude,neilinger2016landau,ivakhnenko2023nonadiabatic}.  

The second mechanism, which is substantially more effective, occurs when an integer multiple of the driving frequency matches the energy difference between two initially separable states. In this case, the concurrence reaches its highest values over  broad regions associated with  ``separable-separable''  resonances (SSR).
Most significantly, we show that by appropriately tuning the  parameters to isolate  a pure SSR  regime, this  dynamically generated entanglement can be precisely controlled through the modulation of the driving amplitude.
To analytically capture this robust entanglement generation we employ Van Vleck (GVV) near-degenerate perturbation theory \cite{kemble1937fundamental, cohen2024atom, hausinger2008dissipative, son2009floquet,hausinger2010dissipative} as a standard rotating-wave approximation (RWA) \cite{oliver2005mach,ivakhnenko2023nonadiabatic, cohen2024atom, son2009floquet} spuriously predicts vanishing transition probabilities for pure SS resonances. 
Within this framework, we demonstrate that the time-averaged concurrence is ruled by an effective, dynamically induced coupling between two dominant entangled Floquet states. We refer to this mechanism as {\it Floquet entanglement generation} (FEG).
However, at specific driving amplitudes, the effective Floquet coupling vanishes exactly, leading to a 
{\it coherent destruction of entanglement} (CDE). 


The paper is organized as follows. 
 In Sec.~\ref{sec: 2}, we introduce the model Hamiltonian and analyze its static energy spectrum. In Sec.~\ref{sec: 3}, we investigate the  population dynamics induced by the parametric driving and identify the distinct, SSR and SER, resonances conditions. In Sec.~\ref{sec: 4}, we  derive an effective Hamiltonian and analyze the dynamics for the  SSR condition employing the Generalized Van Vleck perturbation theory. Within this framework, we discuss the  FEG scenario as well as
the CDE phenomenon connecting both to the intrinsic properties of the Floquet states. We present the conclusions in Sec.~\ref{sec: 5}.  Several   analytical calculations are provided in the Appendices.

\section{Coupled  2-Qubit Model}  
\label{sec: 2}
The system consists of two superconducting qubits coupled through a parametrically driven  harmonic interaction.
Each qubit can be represented  by a two-levels system (TLS) and  the global Hamiltonian written in the computational basis $\lbrace |\uparrow\uparrow\rangle, |\uparrow\downarrow\rangle, |\downarrow\uparrow\rangle, |\downarrow\downarrow\rangle \rbrace$ of  the direct tensor product Hilbert space reads  \cite{wendin2007quantum, kjaergaard2020superconducting,blais2021circuit,rasmussen2021superconducting}:
\begin{equation}
     {H}(t) =  H_0 - \frac{1}{2} \ J(t) \ \sigma_z^{(1)} \otimes\sigma_z^{(2)},
\end{equation}
with
\begin{equation}  
          H_0=   - \frac{1}{2} \sum_{j=1}^2 \left( \epsilon_j \sigma_z^{(j)}  + \Delta_j \sigma_x^{(j)}  \right) - \frac{1}{2} \ J_0 \ \sigma_z^{(1)} \otimes \sigma_z^{(2)}. 
    \label{H_full}  
\end{equation}  
 Here, $H_0$ corresponds to the time independent Hamiltonian of the two qubit system, where each qubit is characterized by  a detuning energy  $\epsilon_j$  and a tunnel splitting  $\Delta_j$. The Pauli matrix operators $\sigma_z^{(j)}$ and \(\sigma_x^{(j)}\)  act on the \(j\)-th qubit and the direct product with the identity matrix of the Hilbert space of the other qubit is understood in the first two terms of $H_0$, but explicitly omitted   to simplify the notation.  The last term in $H_0$ describes the time-independent  longitudinal coupling of strength $J_0$.
The parametric harmonic  driving is defined as:
\begin{equation} 
    J(t) = A \cos(\omega t+\varphi_0),
\end{equation}  
where \(A\) is the driving amplitude, \(\omega\) is the angular frequency, and \(\varphi_0 = \omega t_0\) accounts for the unknown initial phase of the drive. 

We set \(\hbar = 1\), normalize all energy scales by the fixed driving frequency \(\omega\) and  choose the system parameters  \(\epsilon_1 = \epsilon_2 \equiv \epsilon_0\), \(\Delta_1/\omega = 0.1\) and \(\Delta_2/\omega = 0.15\). Throughout this work, we will study the regime $\omega\gg\Delta_{1,2}$ relevant  for small-gap qubits.

\begin{figure}[!b]
    \centering
    \begin{overpic}[width=\columnwidth]{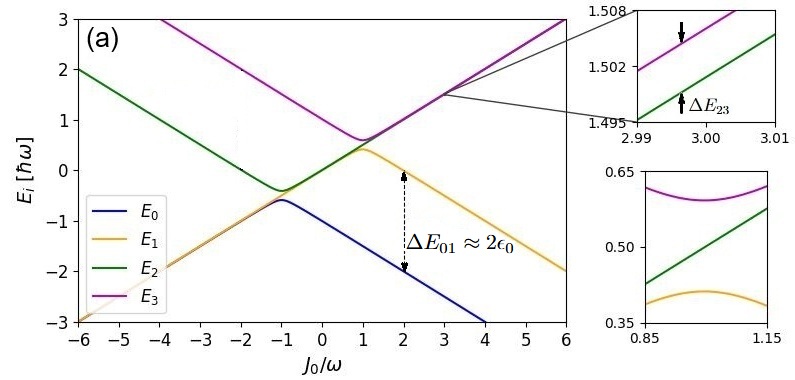}
    \end{overpic}
    
    \vspace{0.1cm}

    \begin{overpic}[width=\columnwidth]{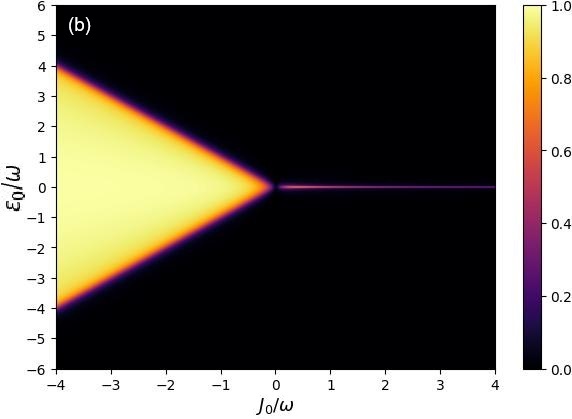}
    \end{overpic}
    \vspace{-0.25cm}
    \caption{(a)Energy spectrum of the static Hamiltonian $H_0$ as function of the normalized static coupling \(J_0/\omega\) for \(\epsilon_0/\omega = 1\).
     The top-right inset shows a zoom of the quasi-degeneracy of \(E_2\) and \(E_3\), while the bottom-right inset highlights  the avoided crossing at \(J_0/\omega = \epsilon_0/\omega\). (b) Color map of the ground state concurrence \(C\), as a function of \(\{\epsilon_0/\omega, J_0/\omega\}\). Two entangled regions can be identified:(i) a triangular region  for \(|J_0/\omega| < |\epsilon_0/\omega|\) in which  \(C_0 \approx 1\) and (ii) a narrow band around \(\epsilon_0/\omega \approx 0\). Outside these regions, the ground state is nearly separable.}
   \label{Fig 1}
\end{figure}

We first analyze the spectrum of \(H_0\), denoting by $E_i$ \(\lbrace i = 0,1,2,3 \rbrace\)  its eigenergies and by  $\{|\phi_i\rangle\}$  the corresponding eigenstates, i.e. $H_0|\phi_i\rangle = E_i|\phi_i\rangle$. The four eigenenergies $E_i$ are displayed  in Fig.~\ref{Fig 1}(a) as a function of the static coupling \(J_0/\omega\) for \(\epsilon_0/\omega = 1\). In the \(\Delta_j = 0\) limit, the Hamiltonian is exactly diagonal in the triplet-singlet basis:\\
\begin{equation}
    \begin{aligned}
    \vert T_+\rangle &= \vert\uparrow\uparrow\rangle, \\
    \vert T_0\rangle &= \frac{1}{\sqrt{2}}\big(\vert\uparrow\downarrow\rangle + \vert\downarrow\uparrow\rangle\big), \\
    \vert T_-\rangle &= \vert\downarrow\downarrow\rangle, \\
    \vert S\rangle &= \frac{1}{\sqrt{2}}\big(\vert\uparrow\downarrow\rangle - \vert\downarrow\uparrow\rangle\big),
\end{aligned}
\end{equation}
and the spectrum consists of two non-degenerate eigenstates, \(\vert T_+\rangle\) and \(\vert T_-\rangle\), and a degenerate subspace spanned by the maximally entangled states \(\vert T_0\rangle\) and \(\vert S\rangle\). In this case, the corresponding eigenenergies exhibit a linear dependence on the static coupling \(J_0\), featuring two exact crossings located at \(J_0 = \pm \epsilon_0\). When a finite \(\Delta_j \neq 0\) is introduced, the degeneracy of the entangled subspace is lifted, and the exact crossings open into avoided crossings as shown in the right panels of Fig.~\ref{Fig 1}(a).

To quantify the degree of entanglement of a pure state \(\vert \Psi \rangle\) across these parameter regimes, we employ the concurrence \cite{wootters1998entanglement}:
\begin{equation}
    C = \Big| \langle \Psi\vert^*\, \sigma_y^{(1)} \otimes \sigma_y^{(2)} \,\vert \Psi \rangle \Big|.  
    \label{eq: C_0}
\end{equation}
This metric ranges from \(0\) for separable states to \(1\) for maximally entangled states.
As shown in Fig.~\ref{Fig 1}(b), the ground state concurrence exhibits a sharp transition at the avoided crossing \(J_0 = -|\epsilon_0|\). Within the region \(J_0 < -|\epsilon_0|\), the ground state exhibits near-maximal entanglement (\(C \approx 1\)).
However, in the region \(J_0 > -|\epsilon_0|\), the ground state becomes nearly separable, with intrinsic entanglement surviving only near \(\epsilon_0 \approx 0\). 
As detailed in the next section, our analysis focuses on the regime where the ground state is separable. In particular, we will focus in the \(J_0 > \vert\epsilon_0\vert\) region, where  the two lowest energy levels, \(E_0 \approx -J_0/2 - \epsilon_0\) and \(E_1 \approx -J_0/2 + \epsilon_0\) correspond, to zero order in $\Delta_j$, to the separable states \(\vert T_+\rangle\) and \( \vert T_- \rangle\), respectively. Conversely, the upper energy levels form a quasi-degenerate pair with  \(E_{2,3} \approx J_0/2\),  and correspond  to the entangled states \(\lbrace\vert T_0 \rangle, \vert S\rangle \rbrace\). The degeneracy is lifted   using second-order perturbation theory in \(\Delta_j\) (see Appendix A for a detailed derivation), which yields:
\begin{equation}
    \Delta E_{23} \equiv E_3-E_2=  \frac{1}{2}\sqrt{ \left( \frac{(\Delta_1^2 - \Delta_2^2)\epsilon_0}{\epsilon_0^2 - J_0^2} \right)^2 + \left( \frac{{2}\Delta_1 \Delta_2 J_0}{\epsilon_0^2 - J_0^2} \right)^2 },
    \label{eq:delta_23}
\end{equation}
as highlighted in the inset of Fig.~\ref{Fig 1}(a). Conversely, in the region \( |J_0|< \vert \epsilon_0\vert\), the entangled character of the excited levels changes: \(\vert \phi_1 \rangle\) and \(\vert \phi_2 \rangle\) become the quasi-degenerate entangled pair, while \(\vert \phi_3 \rangle\) becomes a nearly separable state. Because the physical character of the eigenstates exchanges across these avoided crossings, we will subsequently describe the driven system dynamics by referring to the participating states based strictly on their separable or entangled nature.

\section{DRIVEN DYNAMICS AND ENTANGLEMENT GENERATION}
\label{sec: 3}

In this section we study the generation of entanglement when the coupled qubits are driven by the parametric coupling \(J(t)\). We consider the situation in which the system is initially prepared in the ground state \(|\phi_0\rangle\). 
Since our primary goal is to explore the dynamical generation of entanglement from an initially separable state, we focus our subsequent analysis  in the parameter space \(J_0 > 0\), where the ground state is nearly separable.
The AC drive is  applied at  time \(t_0\), such that \(|\Psi(t_0)\rangle = |\phi_0\rangle\), inducing population transfer between the eigenstates of $H_0$.
\begin{figure}[!t]
    \centering
    \begin{overpic}[width=\columnwidth]{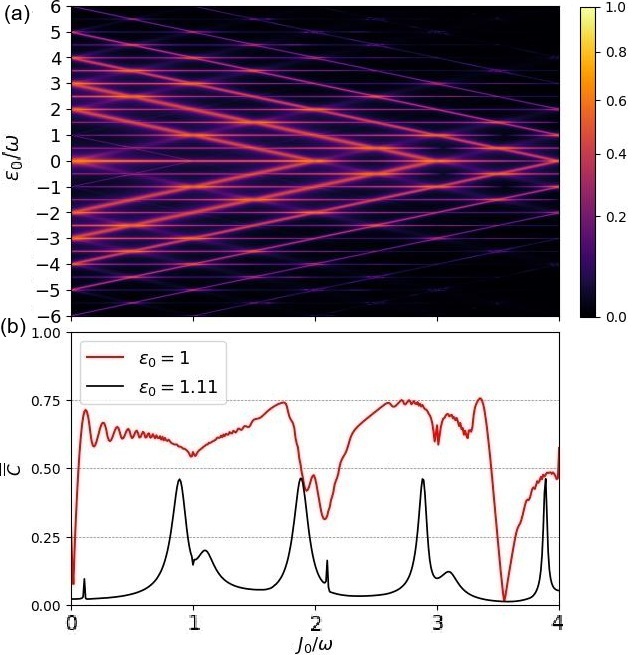}
        \put(390,235){\textcolor{white}{\textbf{(a)}}}
        \put(390,110){\textcolor{black}{\textbf{(b)}}}
    \end{overpic}    
    \vspace{0.25cm}
\caption{(a) Color map of  \(\overline{\overline{C}}\) (Eq.(\ref{eq:Cave})), in the parameter space \(\{\epsilon_0/\omega, J_0/\omega\}\) for \(A/\omega= 3.8\). Patterns where entanglement is strong emerge  within the parameter regime that corresponds to nearly separable region  (\(J_0/\omega > 0\)) in the static case (see Fig.1(b)). (b) \(\overline{\overline{C}}\) for two different resonance conditions. For \(\epsilon_0/\omega = 1.11\)  localized entanglement peaks (\(\overline{\overline{C}} < 0.5\)) appear whenever the system reaches an SER condition. In contrast, for the \(n=2\) SSR condition (\(\epsilon_0/\omega = 1\)) a broad and robust enhancement of entanglement is obtained  with \(\overline{\overline{C}} \approx 0.75\) over a wide range of \(J_0/\omega\). Time evolution was averaged over \(t/T = 700\) and 64 distinct initial phases \(\varphi_0\).}
\label{Fig 2}  
\end{figure} 
\subsection{Time averaged concurrence}

Given an  initial state $|\Psi(t_0)\rangle$ and the evolution operator $U(t,t_0)$,  the wave function at time $t$  is $|\Psi(t)\rangle = U(t, t_0) |\Psi(t_0)\rangle$. Therefore, the concurrence, Eq.(\ref{eq: C_0}), is a two-time quantity $C(t,t_0)$. In order to quantify the typical entanglement dynamically generated for a given  set of parameters of the driving $J(t)$, of period $T=2\pi/\omega$, we evaluate a double time-averaged concurrence,
\begin{equation}\label{eq:Cave}
    \overline{\overline{C}} = \lim_{t' \to \infty} \frac{1}{t'} \int_0^{t'} dt \frac{1}{T} \int_0^T dt_0 C(t, t_0).
\end{equation}
%
This quantity provides a measure of the typical amount of entanglement sustained throughout the duration of the drive, averaged both over the time interval of the drive and  the unknown initial phase \(\varphi_0 = \omega t_0\).
Values of $\overline{\overline{C}}>0.5$ indicate  that the state is  predominantly  entangled on average, whereas  $\overline{\overline{C}}\approx 0$ implies that the state remains mostly separable on (time) average.

In Fig.~\ref{Fig 2}(a) we present the results for the average concurrence, {\it{i.e}},   $\overline{\overline{C}}$, when the initial state of the system is \(|\phi_0\rangle\). A direct comparison with the static case presented in Fig.~\ref{Fig 1}(b) reveals the impact of the parametric drive: in the region \(J_0 > 0\), where  the ground state of the undriven Hamiltonian $H_0$  was nearly  separable  (except for values $\epsilon_0/\omega \sim 0$), the driven system exhibits a robust pattern of lines where  $\overline{\overline C}$ attains values up to 0.75.  A set of  horizontal lines at fixed values of \(2 \epsilon_0 \approx n \omega\), and another set of lines exhibiting a linear dependence on \(J_0/\omega\)  ($\epsilon_{0}  \pm J_0 \approx n\omega $)  are clearly identified. These lines are associated to two set of 
resonances defined respectively  by the condition \( \ \Delta E_{ij}= E_j - E_i \approx n\omega\) (with \(n \in \mathbb{Z}\)) where $j$ and $i$  correspond either to the eigenenergies of  two nearly separable states (SSR)  or to the eigenenergies of a nearly separable and an entangled state (SER), respectively. Taking into account  the energy spectrum    of $H_0$ analyzed in Sec.\ref{sec: 2},  the  two near-resonance conditions are:
\begin{equation}
\begin{aligned}
    2\epsilon_{0} &\approx n\omega \quad \text{(SSR)}, \\
    \epsilon_{0}\pm J_0 &\approx n\omega \quad \text{(SER)}.    
\end{aligned}
    \label{resonance-conditions}
\end{equation}
To illustrate the sensitivity of the driving generated entanglement to each  type of resonance condition,   Fig.~\ref{Fig 2}(b) shows $\overline{\overline{C}}$ as a function of $J_0/\omega$ for two  values of $\epsilon_0/\omega$. For  (\(\epsilon_0/\omega = 1.11\)), the  SER condition is attained for specific values of $J_0/\omega$, resulting in some  defined peaks, but  with values  \(\overline{\overline{C}} < 0.5\). In  contrast, by tuning the system parameters to  the $n=2$  SSR condition (\(\epsilon_0/\omega = 1\)) a robust enhancement of the  entanglement  is obtained with  values of \(\overline{\overline{C}}\approx 0.75\) along a broad region of $J_0/\omega$.

\begin{figure}[b]
    \centering

    \begin{overpic}[width=\columnwidth]{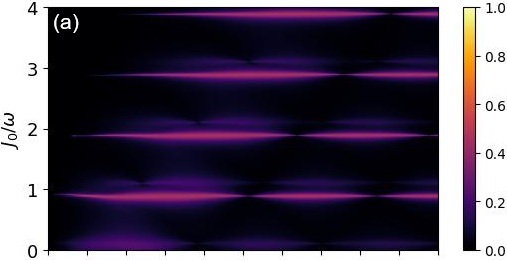}
    \end{overpic}
    
    \vspace{0.1cm}
    
    \begin{overpic}[width=\columnwidth]{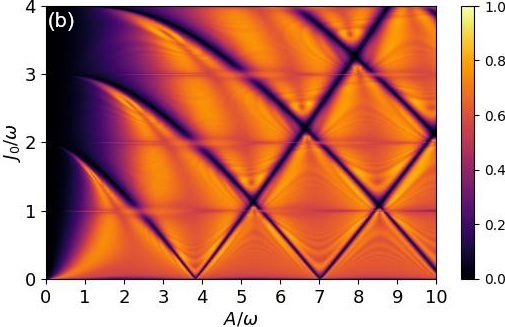}
    \end{overpic}

\caption{ Color map of \(\overline{\overline{C}}\) as a function of \(\{J_0/\omega,\, A/\omega\}\).
(a) For \(\epsilon_0/\omega = 1.11\) resonances where \(\overline{\overline{C}} \approx .5\)  are observed only for values of \(J_0/\omega\) satisfying the SER condition. (b) For the SSR condition \(\epsilon_0/\omega = 1\) (\(n\omega=2\)) the concurrence predominantly exceeds 0.8, exhibiting moderate modulations as a function of the driving amplitude $A/\omega$. See text for more details.} 
    \label{Fig 3}
\end{figure}

To further illustrate the tunability  across the parameter space, we map the double-averaged concurrence as a function of the driving amplitude and the static coupling in Fig.~\ref{Fig 3}. Two qualitatively distinct behaviors emerge depending on the detuning condition. As shown in Fig.~\ref{Fig 3}(a), for the  off-resonant  case (\(\epsilon_0/\omega = 1.11\)), entanglement generation is confined to narrow regions of parameter space associated with the SER condition. Along these resonant lines, the concurrence reaches at most values close to $\overline{\overline{C}}\approx0.5$. Moreover, even within these regions, the entanglement is strongly modulated by the driving amplitude and vanishes for specific values of $A/\omega$.

In contrast,  for $\epsilon_0 /\omega = 1$, corresponding to a pure SSR condition, entanglement is generated over a much broader range of \(J_0\) and \(A\). As displayed in Fig.~\ref{Fig 3}(b), the double-averaged concurrence predominantly exceeds 0.8, exhibiting only moderate modulations as a function of the driving amplitude \(A/\omega\).

The generation of entanglement under SER conditions can be understood as arising from a resonant excitation between the initial separable ground state and an entangled excited state. This process induces Rabi-like oscillations and population transfer between the two resonant states. Consequently, the time-averaged concurrence reaches at most values close to 0.5, since, on average, the system spends only half of the time in the entangled excited state.

In contrast, the entanglement generated under SSR conditions originates from a more intricate mechanism mediated by Floquet states, as described in what follows.

\subsection{Population dynamics}

To understand the generation of entanglement in the SSR case, we start by studying the population transfer between the eigenstates of $H_0$ due to the  drive. We analyze the time-averaged transition probability \(\overline{\overline{P_{ij}}}\)  between eigenstates \(|i\rangle \rightarrow |j\rangle\),
\begin{equation}
    \overline{\overline{P_{ij}}} = \lim_{t' \to \infty} \frac{1}{t'} \int_0^{t'} dt \frac{1}{T} \int_0^T dt_0 P_{ij}(t, t_0),
\end{equation}
which as in the case of $\overline{\overline{C}}$ ,  the average  is performed over  the period $T$ and  the unknown initial phase \(\varphi_0 = \omega t_0\).

\begin{figure}[t]
    \centering
    
    \begin{overpic}[width=\columnwidth]{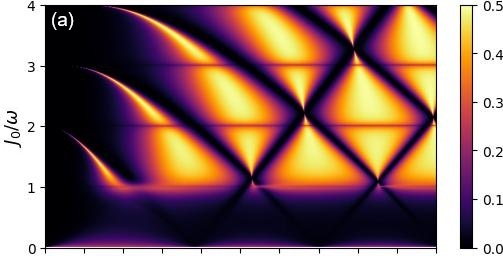}
    \end{overpic}
    
    \vspace{0.2cm}
    
    \begin{overpic}[width=\columnwidth]{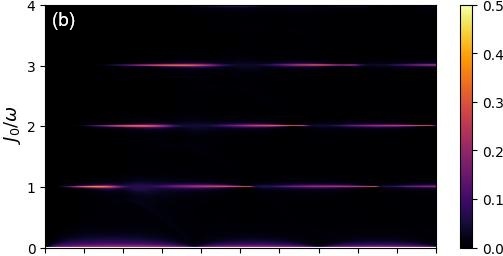}
    \end{overpic}
 
    \vspace{0.2cm}
    
    \begin{overpic}[width=\columnwidth]{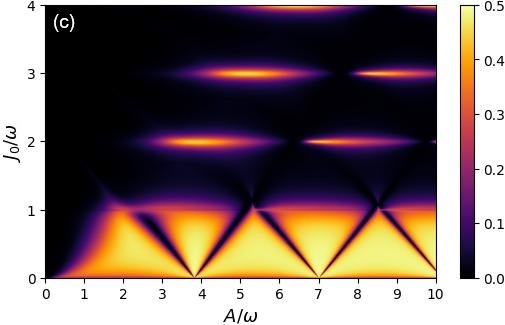}
    \end{overpic}
    
\caption{Double-averaged transition probabilities $\overline{\overline{P_{ij}}}$ in the parameter space $\lbrace J_0/\omega, A/\omega \rbrace$ for  $\epsilon_0/\omega = 1$ (\(n\omega=2\))  SSR condition. A clear change in the plots is observed at  \(J_0/\omega = \epsilon_0/\omega\), corresponding to the location of the avoided in the spectrum of $H_0$ (see  Sec.\ref{sec: 2} for details).  (a) $\overline{\overline{P_{01}}}$ depicts  broad SSR resonances patterns for the region $J_0/\omega > 1$, where $|\phi_1\rangle$  is nearly separable. (b) $\overline{\overline{P_{02}}}$  displays narrow, amplitude-modulated, fringes for values of \(J_0/\omega\) satisfying  the SER condition. In this case $|\phi_2\rangle$ is entangled. (c) $\overline{\overline{P_{03}}}$ displays broad SSR resonances for \(J_0/\omega <1\) ($|\phi_3\rangle$ is nearly separable). Conversely for \(J_0/\omega >1\) ( $|\phi_3\rangle$ is   entangled)  resonances only appear for values of \(J_0/\omega\) satisfying  the SER condition.}
    \label{Fig 4}
\end{figure}

In Fig.~\ref{Fig 4}  the average transition probabilities between the ground  and the  excited states (\(\overline{\overline{P_{01}}},\overline{\overline{P_{02}}},\overline{\overline{P_{03}}}\)) are plotted  as a function of   \(\lbrace J_0/\omega, A/\omega \rbrace\) for $\epsilon_0/\omega=1$. As  we have already  mentioned, the spectrum of $H_0$ exhibits an avoided crossing at \(J_0/\omega = \epsilon_0/\omega\) (see Fig.\ref{Fig 1})  
 and the   separable or entangled character of  the states \(|\phi_1\rangle\) and \(|\phi_3\rangle\) changes depending on whether  \(J_0/\omega>\epsilon_{0}/\omega\) or  \(J_0/\omega<\epsilon_{0}/\omega\), respectively. 
For the region \(J_0/\omega>\epsilon_{0}/\omega\), the state \(|\phi_1\rangle\) is nearly separable. Consequently  \(\overline{\overline{P_{01}}}\)
involves transitions between two states satisfying the SSR condition, leading to a particularly robust generation of entanglement, as shown in Fig.~\ref{Fig 4} (a). In contrast, as  the state \(|\phi_3\rangle\) is entangled in this region, $\overline{\overline{P_{03}}}$ attains appreciable values only along narrow resonant lines corresponding to values of  $J_0$  fulfilling  the SER condition, as observed in  Fig.~\ref{Fig 4} (c). Conversely, for \(J_0/\omega<\epsilon_{0}/\omega\), the state \(|\phi_3\rangle\) is nearly separable and  the state \(|\phi_1\rangle\) is entangled. 
In this case, the SSR condition is satisfied for the states involved in the transition probability \(\overline{\overline{P_{03}}}\) whereas \(\overline{\overline{P_{01}}}\) displays narrow  SER fringes with non-negligible transition probabilities.

Furthermore, since the state \(|\phi_2\rangle\) is  entangled throughout the entire range of   $J_0$ values considered, \(\overline{\overline{P_{02}}}\)  
exhibits narrow resonant features whenever the SER condition is fulfilled. Notice  that  in all these cases the transition
probabilities are strongly modulated by the driving amplitude  $A$, in close analogy with the LSZM interference patterns observed in driven two-level systems as a function of static detuning and driving amplitude \cite{berns2008amplitude,ivakhnenko2023nonadiabatic}.

While the numerical simulations allow us to identify the relevant population transitions associated with the SSR and SER conditions, an analytical description of the amplitude-dependent interference patterns requires a perturbative treatment. We begin by considering the regime where the SSR and SER are simultaneously satisfied, {\it i.e.}  \(\{J_0/\omega = p/2,\, \epsilon_0/\omega = q/2 \mid p,q \in \mathbb{Z},\, (J_0+\epsilon_0)/\omega \in \mathbb{Z}\}\). 
In this regime, and in the limit \(|\phi_0\rangle\approx|T_+\rangle\)and  \(|\phi_1\rangle\approx|T_-\rangle\), we obtain employing  a rotating wave approximation (RWA) \cite{oliver2005mach,ivakhnenko2023nonadiabatic, cohen2024atom, son2009floquet}, the following  analytical expression for the transition probability :
\begin{equation}
\overline{\overline{P_{01}}}=\left|\frac{2\mathcal{J}_{-}\mathcal{J}_{+}\Delta_{1}\Delta_{2}}{\sqrt{[(\Delta_{1}^{2}+\Delta_{2}^{2})(\mathcal{J}_{-}^{2}+\mathcal{J}_{+}^{2})]^{2}-4(\Delta_{1}^{2}-\Delta_{2}^{2})^{2}(\mathcal{J}_{-}\mathcal{J}_{+})^{2}}}\right|^{2}.
\end{equation}
As detailed in Appendix B, this RWA formulation successfully predicts the LZS-like suppression and revival of  the transition probabilities governed by effective energy gaps modulated by Bessel functions of the first kind, \(\mathcal{J}_{\pm}=\mathcal{J}_{-\frac{q }{2} \mp \frac{p}{2}}(A/\omega)\).
However, when the system is tuned solely to an SSR condition, namely when
 \(2\epsilon_{0}=n\omega\) is satisfied but the SER condition is  not (\(\epsilon_{0}\pm J_0\notin\mathbb{Z}\)), the standard RWA  incorrectly predicts a  zero  transition probability \(\overline{\overline{P_{01}}}\) (see Appendix B). This occurs because the purely longitudinal parametric drive cannot directly couple two nearly separable states without virtual transitions through the upper entangled states.

 This failure of the RWA highlights the intrinsically nontrivial nature of entanglement generation in the SSR regime and indicates the necessity of a more refined theoretical treatment, which we address in the following section.

\section{FLOQUET ENTANGLEMENT GENERATION}
\label{sec: 4}

\subsection{Effective Floquet Dynamics Via Van Vleck Perturbation Theory}

The periodicity of the parametric drive enables a standard Floquet treatment \cite{son2009floquet,hausinger2010dissipative}. 
Since the Hamiltonian, Eq.(\ref{H_full}), satisfies \( H(t) = H(t+T) \), Floquet theory ensures solutions of the form \( |\Psi_\nu(t)\rangle = e^{-i \varepsilon_\nu t} |u_\nu(t)\rangle \), where \( \varepsilon_\nu \) denote the quasienergies and \( |u_\nu(t)\rangle \) are the \( T \)-periodic Floquet states.
In this framework, the time-dependent Schrödinger equation maps to a stationary eigenvalue problem \( (H(t) - i \partial_t)|u_\nu\rangle = \varepsilon_\nu |u_\nu\rangle\). 

Since  in our case  \(\Delta_j \ll |J_0|, |\epsilon_0|\), we start by considering the zeroth order limit $\Delta_j=0$, for which the Floquet states are: 
\begin{equation} \label{eq: flst-0-order}
\vert u^{(0)}_{\alpha, n}(t)\rangle
= \vert \alpha \rangle \times e^{in\omega t}\sum_k
\mathcal{J}_{ k}
\!\left( \frac{A}{2\omega} \right) e^{\pm ik\omega t} ,
\end{equation}
where  the sign  \( +\)(\(-\)) is for the Floquet states labeled by 
\(\alpha \in \{T_+, T_-\}\) (\(\alpha \in \{T_0, S\}\)).

The associated zeroth-order quasienergies are given by 
\begin{equation}
\varepsilon^{(0)}_{\alpha,n} = E_\alpha + n\omega,
\end{equation}
where \(\{ E_{i}\}\) are the eigenergies of $H_0$ for $\Delta_j=0$. 
In this limit, the SSR condition is  satisfied when   $\varepsilon^{(0)}_{\uparrow\uparrow, m}=\varepsilon^{(0)}_{\downarrow\downarrow, m-n }$, for $\{m, n \in \mathcal{Z}\}$, {\it i.e.} the two  zeroth-order quasienergies associated to the states  $|T_+\rangle=|\uparrow\uparrow\rangle$ and $|T_-\rangle=|\downarrow\downarrow\rangle$  respectively,  become degenerate.


It turns out (see Appendix C), that this quasienergy degeneracy is lifted only at second order in $\Delta_{1,2}$. To accurately capture the system dynamics under a purely SSR condition, we employ generalized Van Vleck (GVV) near-degenerate perturbation theory. The GVV formalism relies on constructing an effective Hamiltonian, \(H_{\mathrm{GVV}}\), via an unitary transformation that decouples the nearly degenerate Floquet states from the remaining ones. 


As detailed in Appendix C, the effective Hamiltonian in the basis expanded by $\vert u^{(0)}_{\uparrow\uparrow, 0} \rangle$ and $\vert u^{(0)}_{\downarrow\downarrow, -n }\rangle$ and evaluated to second order in \(\Delta_j\) ( \(\Delta_j \ll |J_0|, |\epsilon_0|\)) reduces to:
\begin{equation}
H_{\mathrm{GVV}} = \begin{pmatrix} -\epsilon_0 - \frac{J_0}{2} + \delta_{\uparrow\uparrow} & u \\ u & \epsilon_0 - \frac{J_0}{2} + \delta_{\downarrow\downarrow} - n\omega \end{pmatrix},
    \label{GVV_matrix}
\end{equation}
where the dynamic Stark shifts \(\delta_{\uparrow\uparrow/\downarrow\downarrow}\)  and the effective coupling \(u\) are  given by \cite{son2009floquet,hausinger2008dissipative, hausinger2010dissipative}:
 \begin{equation}
     \begin{aligned}
         \delta_{\uparrow\uparrow/\downarrow\downarrow} = &\mp \frac{(\Delta_1^2 + \Delta_2^2) \pi}{4 \omega\sin((\epsilon_0\pm J_0)\pi/\omega)} \\ 
         &\times \mathcal{J}_{-(\epsilon_0\pm J_0)/\omega} \left(A/\omega \right)\mathcal{J}_{(\epsilon_0\pm J_0)/\omega} \left(A/\omega \right), \\ \\   
         u = & \frac{\Delta_1\Delta_2 \pi \ (-1)^{-2\epsilon_0/\omega}}{2 \omega \sin((\epsilon_0+J_0)\pi/\omega)} \\ 
         &\times\mathcal{J}_{(\epsilon_0-J_0)/\omega} \left(A/\omega \right)\mathcal{J}_{(\epsilon_0+J_0)/\omega} \left(A/\omega \right).\\ 
     \end{aligned}
     \label{GVV-terms}
 \end{equation}
Because \(\delta_{\uparrow\uparrow/\downarrow\downarrow}\) and  \(u\) emerge  as small second-order corrections, the diagonal elements of $H_\mathrm{GVV}$ become nearly identical when  \(n\omega \approx  2\epsilon_0 \), recovering the  SSR condition (see Eq.(\ref{resonance-conditions})). By diagonalizing the effective Hamiltonian Eq.(\ref{GVV_matrix}), evaluated at  the SSR condition, we analytically computed  the quasienergies in second order and the associated Floquet states.

By denoting the obtained quasienergies  $\varepsilon_{\downarrow\downarrow}$ and  $\varepsilon_{\uparrow\uparrow}$, their   splitting results (see Appendix C):
\begin{equation}
    \varepsilon_{\downarrow\downarrow} - \varepsilon_{\uparrow\uparrow} = \sqrt{(\delta_{\uparrow\uparrow} - \delta_{\downarrow\downarrow})^2 + 4 u^2}.
    \label{floquet_gap}
\end{equation}

\begin{figure}[!t]
    \centering
    \begin{overpic}[width=\columnwidth, height=14cm ]{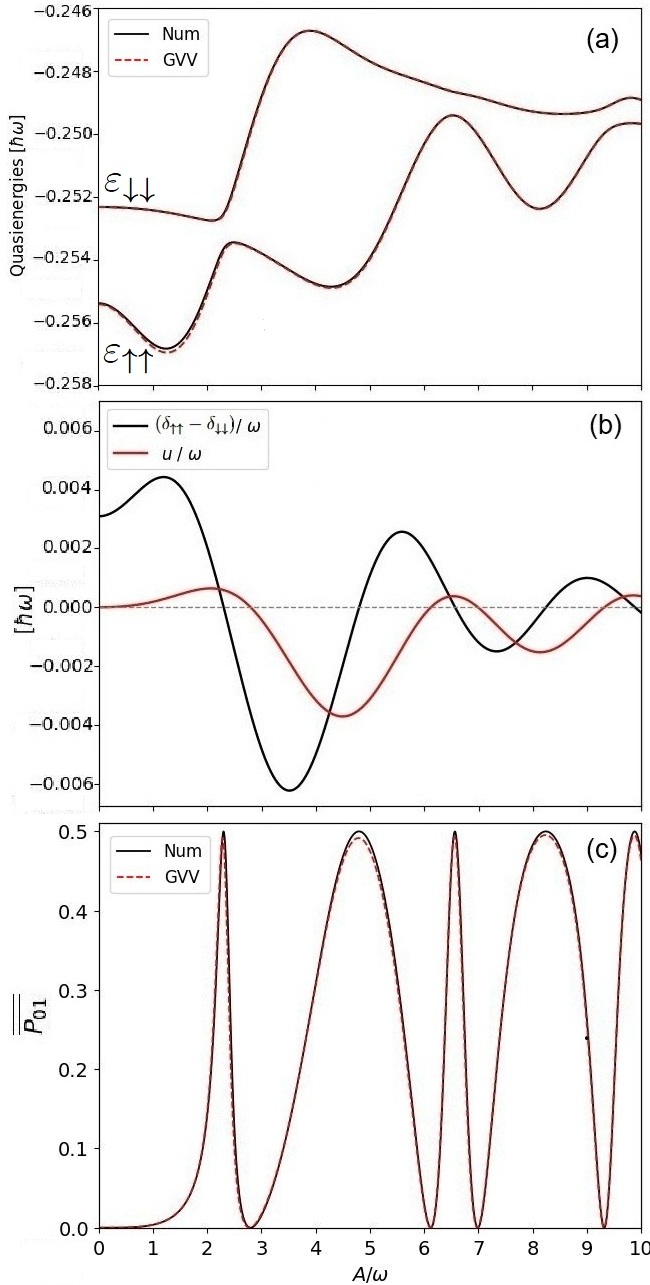}

    \end{overpic}
\caption{Comparison between numerical simulations (solid black lines) and analytical predictions obtained from  the GVV theory (dashed red lines) for a pure SSR condition, \(\epsilon_0/\omega = 1\) and \(J_0/\omega = 2.5\), in all the cases  as a function of the driving amplitude \(A/\omega\). 
(a) Quasienergies  \(\varepsilon_{\uparrow\uparrow}\) and \(\varepsilon_{\downarrow\downarrow}\).  
(b) Effective  coupling \(u\) and dynamic Stark shift difference \(\delta_{\uparrow\uparrow} - \delta_{\downarrow\downarrow}\).  Note that their respective zeros do not coincide. 
(c)Double-averaged transition probability \(\overline{\overline{P_{01}}}\). CDT (\(\overline{\overline{P_{01}}}=0\)) occurs  at the roots of \(u\). 
The maximum population  transfer(\(\overline{\overline{P_{01}}}=0.5\)) is for the condition \(\delta_{\uparrow\uparrow} - \delta_{\downarrow\downarrow} = 0\).}
    \label{Fig 5}
\end{figure}

We demonstrate in Appendix C, through a straightforward calculation,  that the double-averaged transition probability $\overline{\overline{P_{01}}}$ can be written as 
\begin{equation}
        \overline{\overline{P_{01}}} = \frac{2u^2}{(\delta_{\uparrow\uparrow} - \delta_{\downarrow\downarrow})^2 + 4u^2}.
        \label{GVV-prob}
\end{equation}
Note that when only the   SSR is satisfied,  \(\epsilon_0 \pm J \notin \mathbb{Z}\) and  consequently the first kind Bessel functions in  Eq.~\eqref{GVV-terms} are of  non-integer order. In  Figs.~\ref{Fig 5} (a) and  (b)  we compare  the analytical predictions for the quasienergies and for  $\overline{\overline{P_{01}}}$ (obtained using GVV)  with   the corresponding numerical results, showing in both cases an excellent agreement. 
According to Eq.(\ref{GVV-prob}),  the maxima of  $\overline{\overline{P_{01}}}$ as a function of $A/\omega$ are whenever $\delta_{\uparrow\uparrow} - \delta_{\downarrow\downarrow} = 0$, as can be verified from the comparison between  Figs.~\ref{Fig 5} (b) and (c). 

\subsection{Entanglement Generation  from Floquet states}

\begin{figure}[!b]
    \centering
    \begin{overpic}[width=\columnwidth]{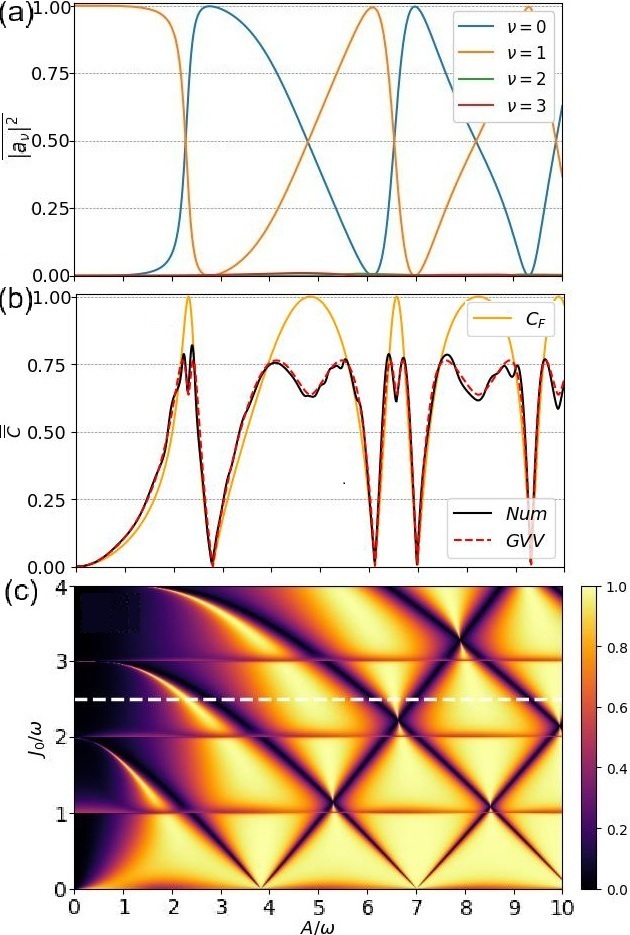}
    \end{overpic}
\caption{(a) Time-averaged squared amplitudes \(\overline{|a_\nu|^2}\)  as a function of \(A/\omega\).
(b) Double-averaged concurrence \(\overline{\overline{C}}\) versus driving amplitude \(A/\omega\). Numerical simulations (solid black line) and the analytical GVV prediction (dashed red line). Floquet concurrence \(C_F\)(solid orange line) is also displayed for comparison.
(c)  \(C_{F}\) as a function of \(J_0/\omega\) and \(A/\omega\) at the SSR condition (\(\epsilon_0/\omega = 1\)). Note the similar pattern with the  double-averaged concurrence \(\overline{\overline{C}}\) shown in Fig.~\ref{Fig 3}(b). All the plots are for the SSR condition (\(\epsilon_0/\omega = 1\), \(J_0/\omega = 2.5\)).}

\label{Fig 6}
\end{figure}

We now demonstrate  how the intrinsic structure of the Floquet states rule the dynamical entanglement generation under the  SSR condition.
As previously discussed, in this case, the parametric drive induces population transfer  between the nearly separable states \(|\phi_0\rangle \approx |T_+\rangle\) and \(|\phi_1\rangle \approx |T_-\rangle\). Consequently, the wavefunction  evolves as a  coherent superposition of these states, dynamically generating an entangled state. To rigorously capture this phenomenon, we expand the time-evolved wave function in the Floquet basis, 
\begin{equation}
    |\Psi(t)\rangle = \sum_\nu a_\nu(t_0) e^{-i\varepsilon_\nu (t-t_0)}|u_\nu(t)\rangle.
    \label{state_vector0}
\end{equation}
As the  initial state  is the nearly separable ground state, \(|\Psi(t_0)\rangle = |\phi_0\rangle\),  the amplitudes  at time $t_0$ are \(a_\nu(t_0) = \langle u_\nu(t_0)|\phi_0\rangle\).
In Fig.~\ref{Fig 6}(a), we plot the $t_0$-averaged  squared amplitudes \(\overline{|a_\nu|^2}\) as a function of $A/\omega$ for a SSR condition $(\epsilon_0/\omega=1)$.
Notice that  throughout  the entire range of values of  $A/\omega$, at most two coefficients remain significant, while all others are negligibly small.
Depending on  $A/\omega$, either \(|a_0|^2\) or \(|a_1|^2\)  or both  have non-vanishing values. Therefore, the state $|\Psi(t)\rangle $  can be written as 
\begin{equation}
|\Psi(t)\rangle \approx a_0(t_0) e^{-i\varepsilon_0 (t-t_0)}|u_0(t)\rangle + a_1(t_0) e^{-i\varepsilon_1 (t-t_0)}|u_1(t)\rangle.
    \label{state_vector}
\end{equation}
and therefore its entanglement is ruled by the intrinsic entanglement of  the two dominant Floquet states $|u_0(t)\rangle$ and $|u_1(t)\rangle$.


As shown in Appendix C, in the GVV approximation, these two dominant Floquet states  take the form:
\begin{equation} \label{eq.Fl_ent}
\begin{aligned}
    |u_0(t)\rangle &\approx \cos\left(\frac{\theta}{2}\right) \vert u^{(0)}_{\uparrow\uparrow}(t) \rangle+ \sin\left(\frac{\theta}{2}\right)e^{-in\omega t}\vert u^{(0)}_{\downarrow\downarrow}(t)\rangle, \\
    |u_1(t)\rangle &\approx -\sin\left(\frac{\theta}{2}\right)\vert u^{(0)}_{\uparrow\uparrow}(t) \rangle + \cos\left(\frac{\theta}{2}\right)e^{-in\omega t}\vert u^{(0)}_{\downarrow\downarrow}(t)\rangle,
\end{aligned}
\end{equation}
where the mixing angle is defined  as \(\tan \theta = 2u / (\delta_{\uparrow\uparrow} - \delta_{\downarrow\downarrow})\). 

The intrinsic concurrence of a given Floquet state is defined as \(C_{F\nu} \equiv |\langle u_\nu(t)^*| \sigma_y^{(1)} \otimes \sigma_y^{(2)} |u_\nu(t)\rangle|\) \cite{sauer_2012}. By evaluating this expression using the states given in Eq.~(\ref{eq.Fl_ent}) we obtain (see Appendix~\ref{appendix: D}) 
\begin{equation}\label{int_co}
     C_{F0} = C_{F1} \approx |\sin \theta|\equiv C_F,
\end{equation}
in terms of the effective mixing angle. 
Numerical results confirm that Eq.(\ref{int_co})
effectively captures the concurrence of the dominant Floquet states. Moreover, by employing Eqs. (\ref{state_vector}) and (\ref{eq.Fl_ent}), within the GVV approximation one can show that the double-averaged concurrence can be expressed approximately as   (see Appendix~\ref{appendix: D} for details of the derivation): 
\begin{equation}
   \overline{\overline{C}} \approx \frac{2}{\pi} \Big[ C_F + (1-C_F^2) \mathrm{arctanh}(C_F) \Big].
\end{equation}

This result demonstrates that the  entanglement observed across the entire \(\{J_0/\omega, A/\omega\}\) parameter space (see Fig.~\ref{Fig 6}(c) and Fig.~\ref{Fig 3}(b)) is ruled by the intrinsic entanglement of the underlying Floquet states. Therefore the  sustained entanglement generated under the SSR condition is a Floquet-induced signature. We named this dynamic phenomenon \textit{Floquet entanglement generation}.

\subsection{Coherent destruction of entanglement}

An important point to notice is that at the precise driving amplitudes where \(u=0\) is ${\overline{\overline{P_{01}}} =0}$, and the mixing angle vanishes (\(\theta \to 0\)). Therefore, the   entanglement of the Floquet states is  destroyed, as can be verified straightforwardly from Eq.(\ref{eq.Fl_ent}). Consequently, the Floquet states revert to uncoupled separable states  yielding to  completely suppressed  double-averaged concurrence $\overline{\overline{C}}$. We term this phenomenon \textit{coherent destruction of entanglement} (CDE). 


As we noted, the CDE mechanism relies on the suppression of a transition probability ${\overline{\overline{P_{01}}}}$, similarly to the well-known coherent destruction of tunneling (CDT) observed in periodically driven two-level systems \cite{grossmann1991coherent, grossmann1992localization, hausinger2010dissipative, kohler2026hidden}. However, unlike CDT which requires an exact quasienergies degeneracy, CDE can occur even in the presence of a finite avoided crossing in the Floquet spectrum.
In fact, the zeros of the effective coupling \(u\) and the difference  \(\delta_{\uparrow\uparrow} - \delta_{\downarrow\downarrow}\)  are governed by distinct mathematical conditions--- the roots of non-integer Bessel functions  and the vanishing of a definite integral Eq.(\ref{zeros_delta}), respectively --- and  therefore  do not in  general coincide, as Fig.~\ref{Fig 5}(c) shows. 
Consequently, the entanglement of the Floquet states can vanish even in the presence of a finite gap in the Floquet spectrum (see Eq.(\ref{floquet_gap})).
By comparing  Fig.~\ref{Fig 5}(a) and Fig.~\ref{Fig 5}(b), one concludes that there are multiple driving amplitudes \(A/\omega\) where \(\overline{\overline{P_{01}}}\) is completely suppressed despite the quasienergies exhibit a finite avoided crossing  (gap). 

The zeros of  the Bessel functions $\mathcal{J}_{(\epsilon_0\pm J_0)/\omega}$   can be analytically approximated when $A/\omega>1$  using the McMahon expansion \cite{abramowitz1972handbook}:
\begin{equation}
    j_{\nu\pm,s} = \left(s + \frac{(\nu\pm)}{\omega} - \frac{1}{4}\right)\pi - \frac{(\nu\pm)^2/\omega^2 - 1/4}{2\pi(s+(\nu\pm)/\omega -1/4)},
    \label{zeros_u}
\end{equation}
where the indices \(\nu\pm = (\epsilon_0\pm J_0)/\omega\) correspond to the respective orders of the two Bessel functions appearing in \(u\) (see Eq.(\ref{GVV-terms})) and \(s\) labels  the   zeros for each order.   
These zeros give the precise amplitudes for which the CDE takes place, in agreement with the numerical results of Fig.~\ref{Fig 6}(b). 
Note  that the dependence on $J_0$, $\epsilon_0$ and $\omega$ provides an effective means to control the coherent generation and destruction of entanglement in the system under study.

\section{Conclusion}
\label{sec: 5}

In conclusion, we have shown that parametrically driven interactions between two qubits provide a novel and highly tunable mechanism for entanglement generation mediated by Floquet states. Besides the conventional Rabi-like population transfer between separable and entangled states associated with SER conditions, we identified a qualitatively distinct regime arising from pure separable–separable resonances (SSR). In this regime, the standard rotating-wave approximation incorrectly predicts vanishing transition probabilities, whereas the full dynamical treatment reveals broad regions of strong entanglement generation.

To analytically capture this phenomenon, we employed generalized Van Vleck (GVV) near-degenerate second order perturbation theory, demonstrating that the time-averaged entanglement is fundamentally inherited from the 
the system's dominant Floquet states. We thus term this mechanism {\it Floquet entanglement generation}. By modulating the external driving amplitude, the generated entanglement can be continuously tuned from strictly zero to highly entangled states.

Our analysis further reveals that the generated entanglement is entirely governed by an effective second-order Floquet coupling parameter  \(u\). Remarkably, this coupling vanishes at specific driving amplitudes corresponding to roots of Bessel functions, leading to a complete dynamical suppression of quantum correlations. We refer  to this effect as \textit {coherent destruction of entanglement} and { unlike the coherent destruction of tunneling, it can  occur even in the presence of a gap in the quasienergies spectrum. } 

The  Floquet entanglement generation discussed in this work could be explored in  several superconducting architectures. In particular, the  model analyzed here may be implemented using fluxonium qubits- which combine  large decoherence times and small leakage outside the computational space- operated detuned away from their sweet spot (such that $\Delta_j\ll \epsilon_0$) and coupled through parametrically modulated capacitive interactions \cite{weiss2022fast, zhang2024tunable, chakraborty2025tunable}.


\section*{Acknowlegments}
We acknowledge support from CNEA and CONICET (PIP 11220220100212CO).

\clearpage


\appendix

\section{Perturbative Calculation of Quasi-Degenerate Energy Levels}
\label{appendix: A}
The energy spectrum analysis presented in Sec.\ref{sec: 2} for the undriven system reveals quasi-degeneracy of the excited states \(\vert\phi_2\rangle\) and \(\vert \phi_3 \rangle\) in the region \(|\epsilon_0|< J_0\).
The undriven system Hamiltonian, Eq.(\ref{H_full}), reads:
\begin{equation}
H_0 = -\frac{1}{2}\sum_{j=1}^2 \left( \epsilon_j \sigma_z^{(j)} + \Delta_j \sigma_x^{(j)} \right) - \frac{1}{2} J_0  \sigma_z^{(1)} \otimes \sigma_z^{(2)}.
\end{equation}
As in our analysis the condition \(\Delta_j \ll |\epsilon_0|, |J_0|\) is satisfied, we can apply perturbation theory to calculate the degeneracy lifting (considering  \(\Delta_{1,2}\) as a perturbation). We thus write:
\begin{equation}
H_0 =  H_z + W,
\end{equation}
with:
\begin{align}
H_z &= -\frac{1}{2}\sum_{j=1}^2 \epsilon_j \sigma_z^{(j)} - \frac{1}{2} J_0 \sigma_z^{(1)} \otimes \sigma_z^{(2)}, \\
W &= -\frac{1}{2}\sum_{j=1}^2 \Delta_j \sigma_x^{(j)}.\\
\end{align}
 \(H_z\) is diagonal in the triplet singlet basis \(\lbrace \vert T_+\rangle, \vert T_0\rangle, \vert T_-\rangle, \vert S\rangle \rbrace\) with degenerate levels  \(E_{2,3}^{(0)} = J_0/2\), corresponding to the zeroth-order eigenstates \(\vert \phi_{2}^{(0)}\rangle = \vert T_0\rangle\) and \(\vert \phi_{3}^{(0)}\rangle = \vert S\rangle\). To lift this degeneracy, we apply degenerate perturbation theory. As the form of \(W\) shows, the degeneracy is not lifted at first order because the projection of \(W\) onto the degenerate subspace \(\mathcal{D}\) vanishes: \(\langle \phi_i^{(0)} \vert W \vert \phi_j^{(0)} \rangle = 0, \forall i,j=2,3\). Therefore, for the second-order correction we compute the matrix with elements
\begin{equation}
\Gamma_{ij} = \sum_{m \notin \mathcal{D}} \frac{\langle \phi_i^{(0)} | W | \phi_m^{(0)} \rangle \langle \phi_m^{(0)} | W | \phi_j ^{(0)}\rangle }{E_{2,3}^{(0)} - E_m^{(0)}}, \hspace{2mm} \lbrace i,j = 2,3 \rbrace.
\end{equation}
 The eigenvalues and eigenvectors of this operator yield the second-order corrections to the eigenenergies and first-order correction to the eigenstates of the system. For our case, evaluating the intermediate states \(m \notin \mathcal{D}\) (which correspond to \(\vert T_+\rangle\) and \(\vert T_-\rangle\)), \(\Gamma_{ij}\) takes the form:
\begin{equation} \label{A7}
    \Gamma = -\frac{1}{{2}(\epsilon_0^2 - J_0^2)}
     \begin{pmatrix}
(\Delta_1 + \Delta_2)^2 J_0 & (\Delta_1^2 - \Delta_2^2)\epsilon_0 \\
(\Delta_1^2 - \Delta_2^2)\epsilon_0 & (\Delta_1 - \Delta_2)^2 J_0 
\end{pmatrix} 
\end{equation}
This \(2 \times 2\) matrix is directly diagonalizable, providing the second-order corrections to the degenerate levels. The eigenvalues of \(\Gamma\) take the form:
\begin{equation}
    \begin{aligned}
            \lambda_{2,3}^{(2)}& =  -\frac{(\Delta_1^2+\Delta_2^2)  J_0}{4 (\epsilon_0^2 - J_0^2)}  \\ 
            &{\mp} \frac{1}{4}\sqrt{ \left( \frac{(\Delta_1^2 - \Delta_2^2) \epsilon_0}{\epsilon_0^2 - J_0^2}\right)^2 + \left(\frac{2 \Delta_1 \Delta_2 J_0}{\epsilon_0^2 - J_0^2} \right)^2}
    \end{aligned}
\end{equation}
The corrected energy levels are given by \(E_{2,3} = J_0/2 + \lambda_{2,3}^{(2)}\).
Also, the corrected eigenvectors can be computed as:
\begin{equation}
\begin{aligned}
        \vert \phi_{2}^{(1)} \rangle &= \frac{\vert T_0\rangle + c_{2} \vert S\rangle }{ \sqrt{ 1 + (c_{2})^2}} \\
            \vert \phi_{3}^{(1)} \rangle &= \frac{\vert S\rangle +d_3\vert T_0\rangle }{ \sqrt{ 1 + (d_{3})^2}}  
\end{aligned}
\end{equation}
with the mixing coefficients defined as \(c_{2} = \frac{\lambda_{2}^{(2)}-\Gamma_{22}}{\Gamma_{23}}\) and \(d_{3} =  \frac{\Gamma_{23}}{\lambda_{3}^{(2)}-\Gamma_{22}}\). 
In the symmetric limit (\(\Delta_1 = \Delta_2\)), the off-diagonal elements of Eq.(\ref{A7}), \(\Gamma_{23} = \Gamma_{32} \propto (\Delta_1^2 - \Delta_2^2)\) vanish, the \(\Gamma\) matrix then becomes strictly diagonal, causing the mixing coefficients asymptotically yield \(c_2 \to 0\) and \(d_3 \to 0\). Consequently, the corrected eigenstates decouple and become the maximally entangled states: \(\vert\phi_2^{(1)}\rangle \to \vert T_0\rangle\) and \(\vert\phi_3^{(1)}\rangle \to \vert S\rangle\).

\section{Transition Probability using RWA for both SSR and SER conditions}
\label{appendix: B}
We obtain an effective description of the system dynamics to calculate the probability of transition between two eigenstates of the system when both SSR and SER conditions are simultaneously satisfied  \(\{J_0/\omega = p/2,\, \epsilon_0/\omega = q/2 \mid p,q \in \mathbb{Z},\, (J_0+\epsilon_0)/\omega \in \mathbb{Z}\}\), using a rotating wave approximation (RWA). To this end, we start with the transformation $\vert \psi'(t) \rangle =  U_\phi \vert \psi(t) \rangle$, $H_\phi = U_\phi H U_\phi^\dagger + i \dot{U_\phi} U_\phi^\dagger$, with $U_\phi$:
\begin{equation}
\begin{aligned}
    U_\phi & = e^{- i \frac{\phi_1}{2} (\sigma_z^{(1)} + \sigma_z^{(2)}) } e^{-i \frac{\phi_2}{2} \sigma_z^{(1)} \sigma_z^{(2)}} \\
    \phi_1 & = \int \epsilon_0 \, dt =  \epsilon_0 t\\
    \phi_2 & = \int (J_0 + A \cos{\omega t}) \, dt = J_0 t + \frac{A}{\omega} \sin{\omega t}
\end{aligned}
\end{equation}
After performing the rotation, we expand the time-dependent  exponentials that will appear in the transformed Hamiltonian in terms of Bessel functions of the first kind:
\begin{equation}
     e^{i(\epsilon_0 \pm J_0)  t} e^{i \frac{A}{\omega} \sin \omega t} =
    \sum_{k = - \infty} ^ {\infty} \mathcal{J}_k \left(\frac{A}{\omega}\right) e^{i(k\omega  + \epsilon_0 \pm J_0 )t }
\end{equation}
Within a rotating-wave approximation, we retain only the time-independent terms in this expansion, considering the double-resonance condition. 
The sum is then  evaluated when \(k = -(\epsilon_0 \pm J_0)/\omega= -(p/2 \pm q/2)\), and the above expression reduces to:
\begin{equation}
    e^{i(\epsilon_0 \pm J_0)  t} e^{i \frac{A}{\omega} \sin \omega t} \simeq \mathcal{J}_{-\frac{q }{2} \mp \frac{p}{2}} \left(\frac{A}{\omega}\right)
\end{equation}
After this approximation, the transformed Hamiltonian  takes the final form:
\begin{equation}
    \begin{aligned}
                H_{\phi}  = &  - \frac{\epsilon_0 -  \frac{q \omega}{2} }{2}(\sigma_z^{(1)} + \sigma_z^{(2)}) \\  &- \frac{J_0 - \frac{p \omega}{2} }{2}  \sigma_z^{(1)} \sigma_z^{(2)} - \frac{1}{2} H''
    \end{aligned}
\end{equation}
where \(H''\) is given by:
\begin{equation}
    H'' = 
 \begin{pmatrix}
0 & \Delta_2^+ & \Delta_1^+ & 0 \\
\Delta_2^+  & 0 & 0 &\Delta_1^- \\
\Delta_1^+& 0 & 0 & \Delta_2^-\\
0 &\Delta_1^- & \Delta_2^- & 0 
\end{pmatrix}   
\end{equation}
Here the couplings \(\Delta^\pm_2 = \Delta_2 \mathcal{J}_{-\frac{q }{2} \mp \frac{p}{2}} \left( \frac{A}{\omega} \right)\) and  \(\Delta_1^\pm = \Delta_1 \mathcal{J}_{-\frac{q }{2} \mp \frac{p}{2}}\left( \frac{A}{\omega} \right)\), are modulated by the first-kind Bessel functions.
Continuing the analysis to calculate the transition probability between two eigenstates of the undriven system $\vert i\rangle \rightarrow \vert f \rangle$ under the transformed Hamiltonian we use the expression:

\begin{equation}
P_{if}(t) = \left\vert \langle f \vert U_\phi(t) e^{-i H_\phi (t-t_0)} U_\phi^\dagger(t_0) \vert i \rangle \right\vert^2,
\end{equation}

We compute the transition probabilities through numerical diagonalization of the transformed Hamiltonian, which yields excellent agreement with full numerical simulations across all transitions. This validates our computational approach for capturing the system's dynamics. For the specific case of the $\vert\phi_0\rangle \to \vert\phi_1\rangle$ transition in the limit where $\vert\phi_0\rangle \approx \vert T_+\rangle$ and $\vert\phi_1\rangle \approx \vert T_-\rangle$, we can further obtain an analytical expression by applying a double averaging procedure:
\begin{equation*}
\overline{\overline{P_{01}}} = \left| 
\textstyle \frac{2 \mathcal{J}_- \mathcal{J}_+ \Delta_1 \Delta_2}
{\sqrt{
\left[ (\Delta_1^2 + \Delta_2^2)(\mathcal{J}_-^2 + \mathcal{J}_+^2) \right]^2 
- 4 (\Delta_1^2 - \Delta_2^2)^2 (\mathcal{J}_- \mathcal{J}_+)^2
}} 
\right|^2
\end{equation*}
\begin{equation}
\mathcal{J}_\pm = \mathcal{J}_{-\frac{q }{2} \mp \frac{p}{2} }\left( \frac{A}{\omega} \right),
\end{equation}

\begin{figure}[!t]
    \centering
    \begin{overpic}[width=\columnwidth]{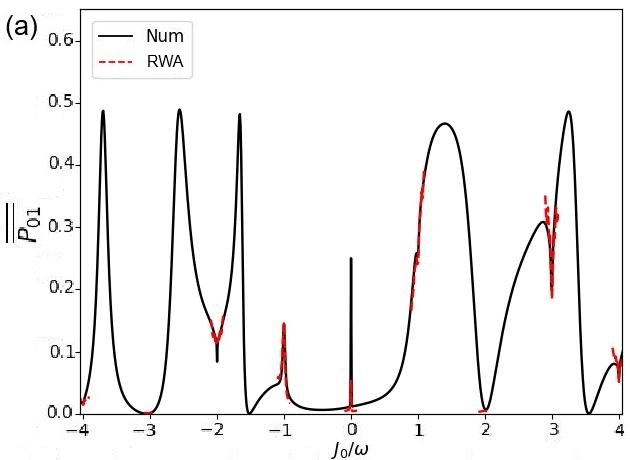}
    \end{overpic}
    
    \vspace{0.3cm} 

    \begin{overpic}[width=\columnwidth]{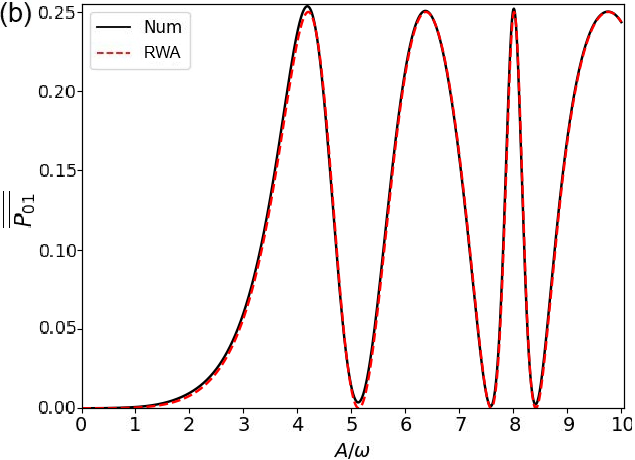}
    \end{overpic}

\caption{
    Transition probability $\overline{\overline{P_{01}}}$ between eigenstates $\vert \phi_0 \rangle \rightarrow \vert \phi_1 \rangle$ for $\epsilon_0/\omega = 1$. Black solid curves show numerical simulations, while red dotted curves represent results from the RWA. (a) Probability as a function of $J_0/\omega$, with the RWA solution shown only in its region of validity. (b) Probability for fixed \(J_0/\omega = 3\) across a range of drive amplitudes \(A/\omega\), demonstrating the regime where the RWA provides an excellent description.}
    \label{Fig 7}
\end{figure}

This result, derived using the RWA under resonance conditions, shows excellent agreement with numerical simulations. Figure~\ref{Fig 7} validates the analytical expression across the parameter space. In particular, panel \ref{Fig 7}(a) demonstrates its accurate description of the $J_0/\omega$ dependence with observed resonances aligning precisely with regions where the RWA remains valid, while panel \ref{Fig 7}(b) shows good agreement between analytical and numerical results as a function of driving amplitude at a fixed coupling strength.

\section{Van Vleck perturbation theory}
\label{appendix: C}
As discussed in the main text, the rotating-wave approximation (RWA) loses its validity in the regime of purely separable-separable resonances (SSR), where the dynamics is governed almost exclusively by transitions between the \(\vert\phi_0 \rangle\) and \(\vert \phi_1 \rangle\) states. In this regime, we show here that the relevant transition probabilities can be obtained analytically using the generalized Van Vleck (GVV) nearly degenerate perturbation theory within the Floquet formalism.

Since the Hamiltonian satisfies \( H(t) = H(t+T) \), Floquet theory ensures solutions of the form \( \vert\Psi_\nu(t)\rangle = e^{-i \varepsilon_\nu t} \vert u_\nu(t)\rangle \), where \( \varepsilon_\nu \) denote the quasienergies and \( \vert u_\nu(t)\rangle \) are the \( T \)-periodic Floquet states. To solve the dynamics, we employ the composite Hilbert space \( \mathcal{H} \otimes \mathcal{T} \) (Sambe space), endowed with the time-averaged inner product \( \langle\langle \cdot \vert \cdot \rangle\rangle = \frac{1}{T} \int_0^T dt \langle \cdot(t) \vert \cdot(t) \rangle \). In this framework, the time-dependent Schrödinger equation maps to a stationary eigenvalue problem \( H_F \vert u_\nu(t)\rangle = \varepsilon_\nu \vert u_\nu(t)\rangle \) governed by the Floquet Hamiltonian \( H_F = H(t) - i \partial_t \). Here it is useful to define the double ket \( \vert u_{\nu, n}\rangle\rangle \), which represents the Floquet states in the Sambe space, where the index \( n \) labels the photon replicas. These states are related to the Fourier coefficients \( \vert u_\nu^{(k)}\rangle \) of the expansion \( \vert u_\nu(t)\rangle = \sum_k e^{ik\omega t} \vert u_\nu^{(k)}\rangle \) by the relation \( \vert u_{\nu, n}\rangle\rangle = \sum_k \vert u_\nu^{(n-k)}\rangle \otimes \vert k) \), where \( \vert k) \) denotes the basis vectors of \( \mathcal{T} \). Defining the unperturbed basis states in the extended space as \( \vert j, k\rangle\rangle \equiv \vert j\rangle \otimes \vert k) \), the long-time double-averaged transition probability between states \( \vert i\rangle \to \vert f\rangle \) is given by \( \overline{\overline{P_{i \to f}}} = \sum_\nu \sum_{k, k'} \vert\langle f \vert u_{\nu}^{(k)}\rangle\vert^2 \vert\langle i \vert u_{\nu}^{(k')}\rangle\vert^2 \). This formulation provides a natural framework for analyzing the physical properties of the system, including the entanglement generation that is the central focus of our investigation.

While Section \ref{sec: 2} utilized the triplet-singlet basis to clarify the physical character of the eigenstates, the application of GVV perturbation theory is significantly streamlined by employing the computational basis \(\{\vert{\uparrow\uparrow}\rangle, \vert{\uparrow\downarrow}\rangle, \vert{\downarrow\uparrow}\rangle, \vert{\downarrow\downarrow}\rangle\}\). In the SSR regime (for \(J_0 > |\epsilon_0| \)), the two states governing the resonant dynamics asymptotically approach the states \(\vert\phi_0\rangle \approx \vert T_+\rangle = \vert{\uparrow\uparrow}\rangle\) and \(\vert\phi_1\rangle \approx \vert T_-\rangle = \vert{\downarrow\downarrow}\rangle\), meaning the computational basis directly encapsulates the resonant subspace. Furthermore, this basis isolates the individual transverse tunneling terms \(\Delta_j\) as off-diagonal perturbations, avoiding the linear combinations (\(\Delta_1 \pm \Delta_2\)) that complicate the notation in the triplet-singlet representation and facilitating a straightforward formulation of the Floquet Hamiltonian. 

Accordingly, the matrix elements of \(H_F\) in the Sambe space are given by:
\begin{equation}
    \langle \langle \alpha, n \vert H_F \vert \beta, m \rangle \rangle = H_{\alpha \beta}^{[n-m]} + n \omega \, \delta_{\alpha \beta} \delta_{nm},
\end{equation}
where \(\alpha,\beta \in \lbrace \uparrow\uparrow, \uparrow\downarrow, \downarrow\uparrow, \downarrow\downarrow \rbrace\), and \(H^{[k]}\) denotes the \(k\)-th Fourier component of the time-dependent Hamiltonian. As a direct consequence of the harmonic functional form of the drive, only three Fourier components are non-vanishing. The static component reads:
\begin{equation}
H^{[0]} =
-\frac{1}{2}
\begin{pmatrix}
    2\epsilon_0 + J_0 & \Delta_2 & \Delta_1 & 0 \\
    \Delta_2 & -J_0 & 0 & \Delta_1 \\
    \Delta_1 & 0 & -J_0 & \Delta_2 \\
    0 & \Delta_1 & \Delta_2 & -2\epsilon_0 + J_0
\end{pmatrix},
\end{equation}
while the components generated by the drive are precisely diagonal:
\begin{equation}
H^{[\pm1]} =
-\frac{1}{4}
\begin{pmatrix}
    A & 0 & 0 & 0 \\
    0 & -A & 0 & 0 \\
    0 & 0 & -A & 0 \\
    0 & 0 & 0 & A
\end{pmatrix}.
\end{equation}
To apply the GVV perturbative method, the Floquet Hamiltonian is decomposed into an unperturbed part \(\mathcal{H}_0\) and a perturbation \(\mathcal{V}\), such that \(H_F = \mathcal{H}_0 + \mathcal{V}\). The perturbation \(\mathcal{V}\) contains the coupling terms \(\Delta_j\). In the Sambe space \(\mathcal{H}_0\) has a block structure, 
\begin{equation}
\begin{aligned}
    (\mathcal{H}_0)_{nn} &= H_0^{[0]} + n\omega \mathbb{I} \\
    &= -\tfrac{1}{2}\,\mathrm{diag}\!\left(2\epsilon_0 + J_0,\,-J_0,\,-J_0,\,-2\epsilon_0 + J_0\right)
    + n\omega \mathbb{I},
\end{aligned}
\end{equation}
where \(n\) denotes the photon index and \(\mathbb{I}\) is the identity matrix in the qubit subspace.
The off-diagonal blocks of \(\mathcal{H}_0\) originate solely from the harmonic modulation of the longitudinal coupling and read
\begin{equation}
    (\mathcal{H}_0)_{nm} = H^{[\pm 1]} \delta_{n,m\pm 1}.
\end{equation}
The perturbation \(\mathcal{V}\) is diagonal in the \textit{n-m} photon index and can be defined using \(V^{[0]}\),
\begin{equation}
\mathcal{V}_{nm} = V^{[0]} \delta_{nm} =
-\tfrac{1}{2}
\begin{pmatrix}
    0 & \Delta_2 & \Delta_1 & 0 \\
    \Delta_2 & 0 & 0 & \Delta_1 \\
    \Delta_1 & 0 & 0 & \Delta_2 \\
    0 & \Delta_1 & \Delta_2 & 0
\end{pmatrix}
\delta_{nm}.
\end{equation}
As a result, all Fourier components of the perturbation vanish except for the static one, i.e., \(V^{[l]} = 0\) for all \(l \neq 0\).\\

Since the unperturbed Hamiltonian \(\mathcal{H}_0\) consists of diagonal operators and longitudinal driving terms, its solutions can be obtained exactly. The corresponding eigenstates admit a Fourier expansion in terms of Bessel functions of the first kind \cite{abramowitz1972handbook},
\begin{equation} 
\vert u^{(0)}_{\alpha, n} \rangle \rangle
= \vert \alpha \rangle \otimes \sum_k
\mathcal{J}_{\eta_\alpha (k-n)}
\!\left( \frac{A}{2\omega} \right)
\vert k \rangle ,
\end{equation}
where \(\eta_\alpha = +1\) for \(\alpha \in \{\uparrow\uparrow, \downarrow\downarrow\}\) and
\(\eta_\alpha = -1\) for \(\alpha \in \{\uparrow\downarrow, \downarrow\uparrow\}\), reflecting the sign of the driving term in \(H^{[\pm1]}\). The associated unperturbed quasienergies are given by the static energies shifted by the photon number,
\begin{equation}
\varepsilon^{(0)}_{\alpha,n} = E_\alpha + n\omega .
\end{equation}
where \(E_\alpha \equiv \langle\alpha\vert H_0^{[0]} \vert\alpha\rangle\) represents the energy of the state \(\vert\alpha\rangle\), distinct from the exact eigenenergies \(E_j\) of the full static system discussed in Section\ref{sec: 2}.
We now construct the Floquet matrix $H_F$ in the basis \(\vert u^{(0)}_{\alpha}, n \rangle\rangle\).
For the diagonal elements it is straightforward 
\(\langle u^{(0)}_{\alpha}, n \vert H_F \vert u^{(0)}_{\alpha}, m \rangle = \varepsilon^{(0)}_{\alpha,n} \delta_{nm}. \)
The off-diagonal matrix elements in this basis, denoted as \(\mathcal{V}'^{[nm]}\), are evaluated using the Bessel-function identity
\( \mathcal{J}_n(u \pm v) = \sum_k \mathcal{J}_{n \mp k}(u)\mathcal{J}_k(v), \)
leading to
\begin{equation}
    \begin{aligned}
        \langle\langle u_{\uparrow \uparrow}^{(0)},n \vert H_F \vert u_{\uparrow \downarrow}^{(0)},m\rangle\rangle &= - \frac{\Delta_2}{2} \mathcal{J}_{n-m} \left( \frac{A}{\omega}\right), \\
        \langle\langle u_{\downarrow \uparrow}^{(0)},n \vert H_F \vert u_{\downarrow \downarrow}^{(0)},m\rangle\rangle &= - \frac{\Delta_2}{2} \mathcal{J}_{-(n-m)} \left( \frac{A}{\omega}\right) ,\\
        \langle\langle u_{\uparrow \uparrow}^{(0)},n \vert H_F \vert u_{\downarrow \uparrow }^{(0)},m\rangle\rangle &= - \frac{\Delta_1}{2} \mathcal{J}_{n-m} \left( \frac{A}{\omega}\right), \\
        \langle\langle u_{\uparrow \downarrow}^{(0)},n \vert H_F \vert u_{\downarrow \downarrow}^{(0)},m\rangle\rangle &= - \frac{\Delta_1}{2} \mathcal{J}_{-(n-m)} \left( \frac{A}{\omega}\right) .
    \end{aligned}
\end{equation}
The remaining matrix elements can be evaluated using the same Bessel-function sum identities and satisfy the symmetry relation \(\langle \langle u^{(0)}_{\beta}, m \vert H_F \vert u^{(0)}_{\alpha}, n \rangle\rangle = (-1)^{n-m} \langle\langle u^{(0)}_{\alpha}, n \vert H_F \vert u^{(0)}_{\beta}, m \rangle\rangle . \)
As a result, the perturbation matrix in the \(\vert u^{(0)}_{\alpha}, n \rangle\rangle\) basis takes the form
\begin{equation}
\mathcal{V'}^{[nm]}= -\frac{1}{2}
\begin{pmatrix}
0 & \Delta_2^+ & \Delta_1^+ & 0 \\
\Delta_2^-  & 0 & 0 & \Delta_1^- \\
\Delta_1^- & 0 & 0 & \Delta_2^- \\
0 & \Delta_1^+ & \Delta_2^+ & 0 
\end{pmatrix},
\end{equation}
where \(\Delta_j^\pm = \Delta_j \mathcal{J}_{n \pm m} \left( \frac{A}{\omega} \right)\).
It is thus evident in this representation that each \(n\)-photon block is coupled to an \(m\)-photon block through the matrix elements \(V'^{[nm]}\). \\

We now apply the GVV nearly degenerate perturbation theory to this problem.
The core of this formalism is to construct an effective Hamiltonian by means of a unitary transformation \(U = e^{iS} \) where \(S\) is a Hermitian operator chosen perturbatively \cite{kemble1937fundamental, cohen2024atom}. The transformed Hamiltonian
\(H_{eff} = e^{iS} H e^{-iS}\) is block diagonal up to a given order in perturbation theory and reproduces, to the same order, the quasienergy spectrum of the original Hamiltonian. Importantly, \(H_{eff}\) only couples states that are nearly degenerate, providing an effective description of the resonant process.

Our goal is to obtain an effective description near the SSR condition,
in the case $J_0>|\epsilon_0|$, for which $|\phi_0\rangle\approx|T_+\rangle=|\uparrow\uparrow\rangle$ and $|\phi_1\rangle\approx|T_-\rangle=|\downarrow\downarrow\rangle$.
In this case,  the SSR condition corresponds to $\varepsilon^{(0)}_{\uparrow\uparrow, 0}=\varepsilon^{(0)}_{\downarrow\downarrow, -n }$.
Then, we focus on the quasi-degenerate Floquet states
$\vert u_{\uparrow \uparrow}, 0 \rangle\!\rangle$ and
$\vert u_{\downarrow \downarrow}, -n \rangle\!\rangle$,
which are resonant at the $n$-photon SSR condition. 
In order to construct the effective Hamiltonian \(H_{eff}\), the generator \(S\) is expanded perturbatively, leading to \( e^{\pm iS} = \mathbb{I} \pm iS^{(1)} \pm iS^{(2)} + \cdots .\) At first order, the nonvanishing matrix elements of the effective Hamiltonian are given by
\begin{equation}
    \begin{aligned}
        \langle\langle u_{\uparrow\uparrow/\downarrow\downarrow}^{(0)}, j \vert H_{eff}^{(1)} \vert u_{\uparrow\uparrow/\downarrow\downarrow}^{(0)}, j \rangle\rangle
        &= \varepsilon^{(0)}_{\uparrow\uparrow/\downarrow\downarrow, j}, \\
        \langle\langle u_{\uparrow\uparrow/\downarrow\downarrow}^{(0)}, 0 \vert H_{eff}^{(1)} \vert u_{\downarrow\downarrow/\uparrow\uparrow}^{(0)}, -n \rangle\rangle
        &= 0 ,
    \end{aligned}
\end{equation}
with \( j \in \lbrace 0, -n \rbrace \). As can be seen, no effective coupling is generated at first order between the near-degenerate states of interest.
Since the results of the generalized Van Vleck method must reproduce those obtained within the rotating-wave approximation (RWA) at leading order \cite{hausinger2010dissipative, son2009floquet}, this explains why the RWA fails to capture the SSR dynamics.
Consequently, a second-order correction is required. The matrix elements of the effective Hamiltonian at this order are calculated as
\begin{equation}
\begin{aligned}
        \langle\langle u^{(0)}_\alpha, n \vert H_{eff}^{(2)} \vert u^{(0)}_{\beta}, m \rangle\rangle& = \frac{1}{2} \sum_{\gamma} \sum_{p}'  \langle\langle u^{(0)}_\alpha, n \vert H_F \vert u^{(0)}_{\gamma}, p \rangle\rangle \\
        &\times \langle\langle u^{(0)}_\gamma, p \vert H_F \vert u^{(0)}_{\beta}, m \rangle\rangle \\
     &\times \left( \frac{1}{\varepsilon^{(0)}_{\alpha, n} - \varepsilon^{(0)}_{\gamma, p}} + \frac{1}{\varepsilon^{(0)}_{\beta, m} - \varepsilon^{(0)}_{\gamma, p}} \right),\\
\end{aligned}
\end{equation}
where the primed summation indicates that intermediate states belonging to the same near-degenerate subspace as the initial and final states are excluded. The matrix elements of the effective Hamiltonian up to second order then acquire the form
\begin{equation}
\begin{aligned}
\langle\langle  u^{(0)}_{\uparrow\uparrow}, 0 \vert &H_{eff}^{(2)} \vert u^{(0)}_{\uparrow\uparrow}, 0 \rangle\rangle  = \\
&-\frac{1}{4} \sum_{m=-\infty}^\infty 
\frac{(\Delta_2^2+\Delta_1^2) \mathcal{J}_{-m}^2 \left( \frac{A}{\omega} \right)}{(\epsilon_0 +J_0)+m\omega}, \\
\langle\langle u^{(0)}_{\downarrow\downarrow}, -n \vert &H_{eff}^{(2)} \vert u^{(0)}_{\downarrow\downarrow}, -n \rangle\rangle 
= \\
&\quad \frac{1}{4} \sum_{m=-\infty}^\infty 
\frac{(\Delta_2^2+\Delta_1^2) \mathcal{J}_{-(n+m)}^2 \left( \frac{A}{\omega} \right)}{(\epsilon_0 -J_0)-(n+m)\omega} ,\\
\langle\langle  u^{(0)}_{\uparrow\uparrow}, 0 \vert &H_{eff}^{(2)} \vert u^{(0)}_{\downarrow\downarrow}, -n \rangle\rangle 
=\\ 
&\frac{1}{4} \sum_{m=-\infty}^\infty 
\Delta_1\Delta_2 \mathcal{J}_{-m}\left( \frac{A}{\omega} \right)\mathcal{J}_{n-m}\left( \frac{A}{\omega} \right)  \\
&\hspace{-11mm} \times 
\left(\frac{1}{(\epsilon_0 -J_0)-(n+m)\omega} - \frac{1}{(\epsilon_0 +J_0)+m\omega} \right).\\
\end{aligned}
\end{equation}
The resulting \(2\times2\) effective Hamiltonian \(H_{GVV}\) up to second order reads
\begin{equation}
    H_{GVV} = 
    \begin{pmatrix}
        -\epsilon_0 - \dfrac{J_0}{2} + \delta_{\uparrow\uparrow} & u\\
        u & \epsilon_0 - \dfrac{J_0}{2} + \delta_{\downarrow\downarrow} - n\omega
    \end{pmatrix},
    \label{H_GVV_app}
\end{equation}
where the diagonal corrections \(\delta_{\uparrow\uparrow} = \langle\langle u^{(0)}_{\uparrow\uparrow}, 0 \vert H_{\mathrm{eff}}^{(2)} \vert u^{(0)}_{\uparrow\uparrow}, 0 \rangle\rangle\) and \(\delta_{\downarrow\downarrow} = \langle\langle u^{(0)}_{\downarrow\downarrow}, -n \vert H_{\mathrm{eff}}^{(2)} \vert u^{(0)}_{\downarrow\downarrow}, -n \rangle\rangle\) account for Stark shifts, while the off-diagonal coupling \(u = \langle\langle u^{(0)}_{\uparrow\uparrow}, 0 \vert H_{\mathrm{eff}}^{(2)} \vert u^{(0)}_{\downarrow\downarrow}, -n \rangle\rangle\) controls the effective hybridization between the unperturbed Floquet states. The SSR condition is given by \(n\omega = 2\epsilon_0\), and the previous expressions can therefore be simplified  using the Newberger sum rule \cite{newberger1982new, russo2024landau}, \(\sum_m (-1)^m\mathcal{J}_{m}(z)\mathcal{J}_{n-m}(z)/(\mu + m) = \pi \mathcal{J}_{\mu}(z)\mathcal{J}_{(n-\mu)}(z)/\sin(\pi\mu)\), with \(\mu \in \mathbb{Q}\). We then obtain,
\begin{equation}
     \begin{aligned}
         \delta_{\uparrow\uparrow} = &- \frac{(\Delta_1^2 + \Delta_2^2) \pi}{4 \omega\sin((\epsilon_0+J_0)\pi/\omega)} \\ 
         &\times \mathcal{J}_{-(\epsilon_0+J_0)/\omega} \left(A/\omega \right)\mathcal{J}_{(\epsilon_0+J_0)/\omega} \left(A/\omega \right), \\ \\
         \delta_{\downarrow\downarrow} = & \frac{(\Delta_1^2 + \Delta_2^2) \pi}{4 \omega \sin((\epsilon_0-J_0)\pi/\omega)}\\
         &\times \mathcal{J}_{-(\epsilon_0-J_0)/\omega} \left(A/\omega \right)\mathcal{J}_{(\epsilon_0-J_0)/\omega} \left(A/\omega \right),\\ \\
         u = & \frac{\Delta_1\Delta_2 \pi \ (-1)^{-2\epsilon_0/\omega}}{2 \omega \sin((\epsilon_0+J_0)\pi/\omega)} \\ 
         &\times\mathcal{J}_{(\epsilon_0-J_0)/\omega} \left(A/\omega \right)\mathcal{J}_{(\epsilon_0+J_0)/\omega} \left(A/\omega \right).\\ 
     \end{aligned}
 \end{equation}
The effective Hamiltonian in Eq.~\ref{H_GVV_app} has the standard form of a two-level system driven by an oscillating field beyond the RWA and admits exact analytical solutions. Its eigenvalues correspond to the second-order quasienergies, while the \textit{n}-photon transition probability \(\overline{P}_{u_{\uparrow\uparrow} \rightarrow u_{\downarrow\downarrow}}^{(n)}\) obtained from this approach coincides with the double-averaged transition probability \(\overline{\overline{P_{01}}}\) in the limit \(\vert\phi_0 \rangle \approx \vert \uparrow\uparrow \rangle\) and \(\vert\phi_1 \rangle \approx \vert \downarrow\downarrow \rangle\). Then both quantities take the form:
\begin{equation}
    \begin{aligned}
        \varepsilon_{\uparrow\uparrow/\downarrow\downarrow} &=
        \frac{1}{2}\left(
        \delta_{\uparrow\uparrow} + \delta_{\downarrow\downarrow}
        \mp \sqrt{(\delta_{\uparrow\uparrow} - \delta_{\downarrow\downarrow})^2 + 4u^2}
        \right), \\
        \overline{\overline{P_{01}}} &=
        \frac{2u^2}{(\delta_{\uparrow\uparrow} - \delta_{\downarrow\downarrow})^2 + 4u^2}.
    \end{aligned}
\end{equation}
As it is evident from the expressions above, the time-averaged transition probability \(\overline{\overline{P_{01}}}\) is maximized when the dynamic Stark shift difference vanishes (\(\delta_{\uparrow\uparrow} - \delta_{\downarrow\downarrow} = 0\)). To analytically determine the driving amplitudes that satisfy this condition, we rewrite this difference using the integral representation of Bessel function products, \(\mathcal{J}_\nu(x)\mathcal{J}_{-\nu}(x) = \frac{2}{\pi} \int_0^{\pi/2} \mathcal{J}_0(2x \cos\Theta) \cos(2\nu\Theta) d\Theta\) \cite{watson1922treatise,russo2024landau}. By applying this identity alongside standard trigonometric relations, the condition for a vanishing Stark shift difference straightforwardly reduces to finding the roots of a single definite integral:
\begin{equation}
    \int_0^{\pi/2} \mathcal{J}_0 \left(\frac{2A}{\omega} \cos \Theta \right) \sin \left( \frac{2\epsilon_0}{\omega} \Theta\right) \sin \left( \frac{2J_0}{\omega} \Theta\right) d\Theta = 0.
    \label{zeros_delta}
\end{equation}

 We conclude this Appendix with a comparison, shown in Fig.~\ref{Fig 5}, of  the GVV results (red dotted lines) with the numerical calculations (solid black lines).  We find that they are in excellent agreement , thereby validating the perturbative Floquet description employed here and confirming the reliability of the GVV approach.
 
 \section{Analytical results for the Concurrence} \label{appendix: D}
As derived in Appendix~\ref{appendix: C}, the first-order Floquet states obtained by diagonalizing the effective Hamiltonian \(H_{GVV}\) (Eq.~\ref{H_GVV_app}) can be used to analytically evaluate the  concurrence. By expanding the time-evolved wave function in the Floquet basis (Eq.~\ref{state_vector}), the time-dependent concurrence is given by:
\begin{equation} \label{ctto}
    C(t,t_0) = \left| \sum_{\alpha,\beta} a_\alpha(t_0) a_\beta(t_0) e^{-i(\varepsilon_\alpha+\varepsilon_\beta)(t-t_0)} \tilde{C}_{\alpha\beta}(t) \right|,
\end{equation}
where
\begin{equation}
    \begin{aligned}
         a_\alpha(t_0) &= \langle u_\alpha(t_0)|\Psi(t_0)\rangle,\\
         \tilde{C}_{\alpha, \beta}(t) &= \langle u_\alpha(t)|^* \sigma_y^{(1)} \otimes \sigma_y^{(2)}| u_\beta (t) \rangle.
    \end{aligned}
\end{equation}
In the regime of interest, \(a_\nu(t_0) \neq 0\) only for \(\nu \in \{0, 1\}\). Thus, the concurrence simplifies to:
\begin{equation} \label{D3}
\begin{aligned}
    C(t,t_0) =  \Big| 2a_0(t_0) a_1(t_0) \tilde{C}_{01}(t)& + a_0(t_0)^2 \tilde{C}_{00}(t) e^{i\Delta\varepsilon(t-t_0)} \\
    &+ a_1(t_0)^2 \tilde{C}_{11}(t)  e^{-i\Delta\varepsilon(t-t_0)} 
      \Big|,
\end{aligned}
\end{equation}
 {where  we have used  that $\tilde C_{01}=\tilde C_{10}$}
and \(\Delta\varepsilon = \varepsilon_1 - \varepsilon_0\). Since the Floquet concurrence is defined as \(C_{F\nu} \equiv |\tilde{C}_{\nu\nu}|\), the total time-dependent entanglement is governed by the concurrence of the dominant Floquet states and an interference term {containing} \(\tilde{C}_{01}(t)\). Evaluating these quantities using the explicit form of the states within the GVV approximation (Eqs.~(\ref{eq: flst-0-order}) and (\ref{eq.Fl_ent})) yields:
\begin{equation} \label{D4}
    \begin{aligned}
        a_{0}(t_0) &=  \cos(\theta/2)   e^{-i \xi(t_0)},\\
        a_{1}(t_0) &=  - \sin(\theta/2) e^{-i \xi(t_0)},\\
        \tilde{C}_{00}(t) = -\tilde{C}_{11}(t) &= -\sin(\theta) e^{-i \chi(t)},\\
        \tilde{C}_{01}(t) = \tilde{C}_{10}(t) &= -\cos(\theta) e^{-i\chi(t)},
    \end{aligned}
\end{equation}
where we have defined the time-dependent phases \(\xi(t_0) = \frac{A}{2\omega} \cos(\omega t_0) - n\omega t_0\) and \(\chi(t) = \frac{A}{\omega} \cos(\omega t) - 3n\omega t\). From these expressions, it is straightforward to verify that \(C_F \equiv C_{F0} = C_{F1} = |\sin \theta|\). 

Substituting Eq.(\ref{D4}) into Eq.~(\ref{D3}), the time-dependent phases completely factor out of the absolute value. 
By applying standard trigonometric identities, the time-dependent concurrence can be rewritten as:
\begin{equation}
\begin{aligned}
    C(t,t_0) &= 2 \left|\sin\theta \sin\left( \frac{(t-t_0)\Delta\varepsilon}{2}\right)\right|\\ &\times \sqrt{1 - \sin^2\theta \sin^2\left( \frac{(t-t_0) \Delta\varepsilon}{2}\right) }.
\end{aligned}
\end{equation}
We now calculate the double-averaged concurrence \(\overline{\overline{C}}\). We note that \(C(t,t_0)\) depends on time exclusively through the dimensionless variable \(x = (t-t_0) \Delta\varepsilon/2\), and is periodic in \(x\) with a period of \(\pi\). Therefore:
\begin{equation}
    \begin{aligned}
        \overline{\overline{C}} &= \lim_{t' \to \infty} \frac{1}{t'} \int_0^{t'} dt \frac{1}{T} \int_0^T dt_0 \, C(t, t_0) \\
        &= \frac{1}{T} \int_0^T dt_0 \left[ \lim_{t' \to \infty}  \frac{1}{t'} \int_0^{t'} dt \, C(t- t_0) \right] \\
        &= \frac{1}{T} \int_0^T dt_0 \left[ \frac{1}{\pi} \int_0^{\pi} C(x) \, dx \right]. 
    \end{aligned}
\end{equation}
Since the term in brackets is strictly independent of \(t_0\), the outer average becomes trivial. Consequently, the double average reduces precisely to a single integral over one period of \(x\):
\begin{equation}
    \overline{\overline{C}} = \frac{1}{\pi}\int_0^\pi 2\vert\sin\theta \sin x\vert \sqrt{1 - \sin^2\theta \sin^2x} \, dx.
\end{equation}
Furthermore, the integrand is completely symmetric around \(x=\pi/2\), then we can restrict the integration domain to \([0, \pi/2]\) ( multiplying by a factor 2). Over this restricted interval, \(\sin x\) is strictly non-negative, allowing us to drop the absolute value. Applying the substitution \(u = \cos x\), the integral straightforwardly transforms to:
\begin{equation}
    \overline{\overline{C}} = \frac{4|\sin \theta|}{\pi} \int_0^1 \sqrt{\cos^2\theta + \sin^2\theta \, u^2} \ du.
\end{equation}
Evaluating this integral yields:
\begin{equation}
    \overline{\overline{C}} = \frac{2 |\sin\theta|}{\pi} + \frac{2(1-\sin^2\theta)}{\pi} \mathrm{arcsinh}(|\tan \theta|).
\end{equation}
Finally, one can express this result entirely in terms of the intrinsic Floquet concurrence \(C_F = |\sin \theta|\). By applying the  identity \(\mathrm{arcsinh}(|\tan\theta|) = \mathrm{arctanh}(|\sin\theta|) = \mathrm{arctanh}(C_F)\), we obtain, within the GVV approximation, the closed-form expression:
\begin{equation} \label{D9}
    \overline{\overline{C}} = \frac{2}{\pi} \Big[ C_F + (1-C_F^2) \mathrm{arctanh}(C_F) \Big].
\end{equation}
We further analyze the asymptotic limits of \(\overline{\overline{C}}\), starting with \(\theta = \pi/2\) (or \(C_F \to 1\)). In this case it is straightforward to show that the second term of Eq. (\ref{D9}) vanishes,
 and the time-averaged concurrence attains  the suboptimal value of \(\overline{\overline{C}} \to 2/\pi \approx 0.636\), which corresponds to the local minima (valleys in Fig.6(b)) of \(\overline{\overline{C}}\). 
This suppression can be understood using Eq.~(\ref{D3}). At \(\theta = \pi/2\) the dominant Floquet states are equally populated (\(|a_0(t_0)|^2 = |a_1(t_0)|^2 = 1/2\)), and the interference term vanishes,  (\(\tilde{C}_{01} \propto \cos(\pi/2) = 0\)). 


To determine the absolute maximum, we set \(d\overline{\overline{C}}/dC_F = 0\), which leads to the transcendental equation \(C{_F}  \, \mathrm{arctanh}(C_F) = 1\) with root \(C_F \approx 0.834\) (corresponding to \(\theta \approx 0.98 + l\pi\) and \(\theta \approx 2.16 + l\pi\), with \(l \in \mathbb{Z}\)). Evaluating Eq.(\ref{D9}) for this value of $\theta$ we get \(\overline{\overline{C}}_{\text{max}} \approx 0.763\). \\
Physically, at this point, the populations of the Floquet states become unbalanced and the  interference term in Eq.~(\ref{D3}) becomes non-zero (\(\tilde{C}_{01} \neq 0\)). This  damps the coherent oscillations obtained  for \(\theta = \pi/2\). Consequently, while the concurrence of the Floquet states is no longer maximal, the time-dependent concurrence is prevented from dropping to zero. This mitigation of complete destructive interference is what ultimately maximizes the time-averaged entanglement.

Finally, we analyze the limit, corresponding to the vicinity of Coherent Destruction of Entanglement (CDE). In this regime, the effective Floquet coupling approaches zero (\(\theta \to 0\))
. Performing a Taylor expansion on Eq.~(\ref{D9}) and retaining only the leading order yields \(\overline{\overline{C}} \approx \frac{4}{\pi} C_F\).\\
This behavior is naturally recovered from Eq.~(\ref{D3}). As \(\theta \to 0\), the system is predominantly localized in the initial separable state (\(|a_0|^2 \approx 1\), \(|a_1|^2 \approx 0\)). 
Consequently, both the interference term and the coherent oscillation amplitude are severely suppressed. The concurrence then scales linearly with the minimal entanglement \(C_F\) surviving in the Floquet basis.

\newpage

\bibliography{bibliography}
\end{document}